\documentclass[12pt]{article}

\usepackage{axodraw4j}
\usepackage{pstricks}
\usepackage{color}

\usepackage{cite}
\usepackage{epsfig}
\usepackage{amssymb}
\makeatletter

\def\paragraph{\@startsection{paragraph}{4}{\z@}{+2.00ex plus
 +1ex minus +.2ex}{1.5ex plus .2ex}{\it\normalsize}}

\def\section{\@startsection {section}{1}{\z@}{+3.0ex plus +1ex minus
  +.2ex}{2.3ex plus .2ex}{\normalsize\bf\boldmath}}
\def\subsection{\@startsection{subsection}{2}{\z@}{+2.5ex plus +1ex
minus +.2ex}{1.5ex plus .2ex}{\normalsize\bf\boldmath}}
\def\subsubsection{\@startsection{subsubsection}{3}{\z@}{+3.25ex plus
 +1ex minus +.2ex}{1.5ex plus .2ex}{\normalsize\it}}

 \expandafter\ifx\csname mathrm\endcsname\relax\def\mathrm#1{{\rm #1}}\fi

\newcounter{saveeqn}

\@addtoreset{equation}{section}

\def\asymp#1%
{\mathrel{\raisebox{-.4em}{$\widetilde{\scriptstyle #1}$}}}

\def\Nequal#1%
{\mathrel{\raisebox{-.5em}{$\stackrel{=}{\scriptstyle\rm#1}$}}}
\newcommand{\dsl}[1]{\not \hspace{-0.7mm}#1}
\def\dsl{\mathpalette\make@slash}
\def\make@slash#1#2{\setbox\z@\hbox{$#1#2$}%
  \hbox to 0pt{\hss$#1/$\hss\kern-\wd0}\box0}

\def\refeq#1{\mbox{(\ref{#1})}}
\def\refeqs#1{\mbox{(\ref{#1})}}

\def\reffi#1{\mbox{Figure~\ref{#1}}}
\def\reffis#1#2{\mbox{Figures~\ref{#1} and \ref{#2}}}
\def\refta#1{\mbox{Table~\ref{#1}}}

\def\refse#1{\mbox{Section~\ref{#1}}}

\def\citere#1{\mbox{Ref.~\cite{#1}}}
\def\citeres#1{\mbox{Refs.~\cite{#1}}}

\newcommand{\TeV}{\unskip\,\mathrm{TeV}}
\newcommand{\GeV}{\unskip\,\mathrm{GeV}}

\newcommand{\fb}{\unskip\,\mathrm{fb}}

\def\mathswitchr#1{\relax\ifmmode{\mathrm{#1}}\else$\mathrm{#1}$\fi}

\newcommand{\PV}{V}
\newcommand{\PW}{\mathswitchr W}

\newcommand{\PZ}{\mathswitchr Z}

\newcommand{\Pg}{\mathswitchr g}

\newcommand{\PH}{\mathswitchr H}

\newcommand{\Pne}{\mathswitch \nu_{\mathrm{e}}}

\newcommand{\Pnmu}{\mathswitch \nu_{\mu}}
\newcommand{\Pd}{\mathswitchr d}

\newcommand{\Pu}{\mathswitchr u}

\newcommand{\Ps}{\mathswitchr s}

\newcommand{\Pc}{\mathswitchr c}

\newcommand{\Pb}{\mathswitchr b}

\newcommand{\Pp}{\mathswitchr p}
\newcommand{\Pt}{\mathswitchr t}

\newcommand{\Pep}{\mathswitchr {e^+}}

\newcommand{\Pmup}{\mathswitchr {\mu^+}}
\newcommand{\Pmum}{\mathswitchr {\mu^-}}

\newcommand{\PWp}{\mathswitchr {W^+}}
\newcommand{\PWm}{\mathswitchr {W^-}}

\def\mathswitch#1{\relax\ifmmode#1\else$#1$\fi}

\newcommand{\Mt}{\mathswitch {m_\Pt}}

\newcommand{\rw}{\mathswitchr w}

\newcommand{\GF}{\mathswitch {G_\mu}}

\newcommand{\alphas}{\alpha_\mathrm{s}}

\def\lra{\mathop{\mathrm{\leftrightarrow}}\nolimits}

\def\draftdate{\relax}
\def\mda{\relax}
\def\mua{\relax}
\def\mla{\relax}
\def\Mda{\relax}
\def\Mua{\relax}
\def\Mla{\relax}
\def\draft{
\def\thtystars{******************************}
\def\sixtystars{\thtystars\thtystars}
\typeout{}
\typeout{\sixtystars**}
\typeout{* Draft mode!
         For final version remove \protect\draft\space in source file *}
\typeout{\sixtystars**}
\typeout{}
\def\draftdate{\today}
\def\mua{\marginpar[\boldmath\hfil$\uparrow$]%
                   {\boldmath$\uparrow$\hfil}%
                    \typeout{marginpar: $\uparrow$}\ignorespaces}
\def\mda{\marginpar[\boldmath\hfil$\downarrow$]%
                   {\boldmath$\downarrow$\hfil}%
                    \typeout{marginpar: $\downarrow$}\ignorespaces}
\def\mla{\marginpar[\boldmath\hfil$\rightarrow$]%
                   {\boldmath$\leftarrow $\hfil}%
                    \typeout{marginpar: $\lra$}\ignorespaces}
\def\Mua{\marginpar[\boldmath\hfil$\Uparrow$]%
                   {\boldmath$\Uparrow$\hfil}%
                    \typeout{marginpar: $\uparrow$}\ignorespaces}
\def\Mda{\marginpar[\boldmath\hfil$\Downarrow$]%
                   {\boldmath$\Downarrow$\hfil}%
                    \typeout{marginpar: $\downarrow$}\ignorespaces}
\def\Mla{\marginpar[\boldmath\hfil$\Rightarrow$]%
                   {\boldmath$\Leftarrow $\hfil}%
                    \typeout{marginpar: $\lra$}\ignorespaces}
\overfullrule 5pt
\oddsidemargin -15mm
\oddsidemargin -10mm
\marginparwidth 29mm
}

\def\stars{\strut\leaders\hbox{*}\hfill\strut}
\def\starline{\hfil\strut\hfil\hbox to \textwidth {\stars}\hfil}

\usepackage[hang, small, bf, margin=20pt, tableposition=top]{caption}
\usepackage{epsfig,graphicx}
\usepackage{setspace}
\usepackage{tabularx}
\usepackage{chngpage}
\usepackage{booktabs}
\usepackage{pstricks}
\usepackage{color}

\usepackage{float}
\usepackage{subfig}
\usepackage{slashed}
\usepackage{amsmath}
\usepackage{appendix}

\usepackage{sectsty}
\usepackage{hyperref}
\chapterfont{\large\sc\centering}
\chaptertitlefont{\centering}
\subsectionfont{\centering}

\newcommand{\beq}{\begin{equation}}
\newcommand{\eeq}{\end{equation}}

\newcommand{\beqa}{\begin{eqnarray}}
\newcommand{\eeqa}{\end{eqnarray}}

\newcommand{\ba}{\begin{align}}
\newcommand{\ea}{\end{align}}

\newcommand{\bit}{\begin{itemize}}
\newcommand{\eit}{\end{itemize}}

\newcommand{\f}{\frac}
\newcommand{\nn}{\nonumber}
\newcommand{\mc}[3]{\multicolumn{#1}{#2}{#3}}

\newcommand{\rW}{\mathrm{W}}
\newcommand{\rZ}{\mathrm{Z}}
\newcommand{\ri}{\mathrm{i}}
\newcommand{\re}{\mathrm{e}}
\newcommand{\rj}{\mathrm{j}}
\newcommand{\rd}{\mathrm{d}}

\newcommand{\ru}{\mathrm{u}}
\newcommand{\rc}{\mathrm{c}}
\newcommand{\rp}{\mathrm{p}}
\newcommand{\rs}{\mathrm{s}}

\newcommand{\muR}{\mu_{\mathrm R}}
\newcommand{\muF}{\mu_{\mathrm F}}
\newcommand{\pT}{p_{\mathrm T}}
\newcommand{\rT}{\mathrm T}

\begin{document}
\thispagestyle{empty}
\def\thefootnote{\fnsymbol{footnote}}
\setcounter{footnote}{1}
\null
\draftdate
\strut\hfill {ZU-TH 19/12}\\
\strut\hfill {LPN12-093}\\
\vspace{1.5cm}
\begin{center}
{\Large \bf\boldmath
{NLO QCD corrections to $\PWp\PWp \rj\rj$ production in vector-boson
  fusion at the LHC}
\par}
\vspace{1.5cm}
{\large
{\sc A.\ Denner$^1$, L. Ho\v sekov\'a$^{2}$ and S.~Kallweit$^3$} } \\[.5cm]
$^1$ {\it Universit\"at W\"urzburg, Institut f\"ur Theoretische Physik und Astrophysik,\\
D-97074 W\"urzburg, Germany}
\\[0.5cm]
$^2$ {\it Instituto de F\'isica Corpuscular,\\
UVEG - Consejo Superior de Investigaciones Cient\'ificas,\\
E-46980 Paterna (Valencia), Spain%
}\\[0.5cm]
$^3$ {\it Institut f\"ur Theoretische Physik, Universit\"at Z\"urich,\\
CH-8057 Z\"urich, Switzerland}

\par \vskip 1em
\end{center}\par
\vfill {\bf Abstract:}
\par
We present a next-to-leading-order QCD calculation for
\mbox{$\Pep\Pne\Pmup{\nu}_\mu jj$} production in vector-boson fusion,
i.e.\ the scattering of two positively charged $\PW$ bosons at the
LHC. We include the complete set of electroweak leading-order diagrams
for the six-particle final state and quantitatively assess the size
of the $s$-channel and interference contributions in VBF kinematics.
The calculation uses the complex-mass scheme to describe the W-boson
resonances and is implemented into a flexible Monte Carlo generator.
Using a dynamical scale based on the transverse momenta of the
jets, the QCD corrections stay below about $10\%$ for all considered
observables, while the residual scale dependence is at the level
of~$1\%$.
\par
\vskip 2.5cm
\noindent
September 2012
\null
\setcounter{page}{0}
\clearpage
\def\thefootnote{\arabic{footnote}}
\setcounter{footnote}{0}

\section{Introduction}
Vector-boson fusion (VBF) processes at the Large Hadron Collider (LHC)
offer unique signatures owing to two easily identifiable forward and
backward jets. This process class is not only useful for confirming
the existence of the Standard Model Higgs boson, but in particular for
studying its characteristics, including its couplings to both fermions
and electroweak (EW) vector bosons~\cite{Duhrssen:2004cv,Belyaev:2002ua} 
and its CP properties~\cite{Hankele:2006ma,Hagiwara:2009wt}.

VBF processes involving the scattering of vector bosons constitute an
irreducible background to Higgs-boson production in association with
two jets, in particular for \mbox{$\PH\to \PZ\PZ/\PWp\PWm\to4l$} decay
modes as they share the same final states. It is therefore desirable
to obtain accurate theoretical predictions and error estimates for
these background processes.  The reactions of the type
\mbox{$\Pp\Pp\to \PV\PV\rj\rj\to 4l\rj\rj+X$} are also seen as an
important probe of the EW symmetry breaking
itself~\cite{Bagger:1995mk}.  Without the presence of the Higgs boson
perturbative unitarity of the Standard Model at very high energy
scales would be violated in processes involving weak-vector-boson
scattering unless some other mechanism beyond that described by the
Standard Model controls the unphysical behaviour (see e.g.\ 
\citere{Ballestrero:2010vp}).  Moreover, VBF into pairs of vector
bosons is an important background to various searches for new physics.

We are specifically interested in VBF processes that involve the
scattering of weak gauge bosons and lead to final states with two
jets and four leptons (charged leptons and neutrinos).
At leading order (LO), two hard production mechanisms give rise to
these final states. The purely EW
contributions of order $\alpha^6$ involve in particular the genuine
VBF contributions, i.e.\ diagrams where vector bosons are emitted from
the incoming \mbox{(anti-)quarks}, then scatter and decay into pairs
of leptons.  In addition, there are QCD-production contributions of
order $\alpha^4\alpha_\rs^2$ which proceed via gluon-mediated
\mbox{(anti-)quark} scattering processes or processes with two
external gluons and two external \mbox{(anti-)quarks}, where in
both cases the two EW vector bosons are
emitted from the \mbox{(anti-)quark} line(s).
Since the jets in the QCD production mode tend to be closer
in rapidity than in the EW production mode, a cut requiring a large
rapidity separation between these jets or more generally a central jet
veto suppresses the QCD production mode by two orders of magnitude as
demonstrated in
\citeres{Rainwater:1996ud,Jager:2011ms,Ballestrero:2010vp}. Owing to
the different colour structure of QCD and EW production modes,
interference terms between these mechanisms are doubly suppressed at
LO; they only appear at sub-leading colour and if all quarks are
identical. In our calculation, we restrict ourselves to the EW
production mode.

In this paper we focus on the process involving two jets, two
positively charged leptons and two neutrinos in the final state, i.e.\
\mbox{$\Pp\Pp\to \PWp\PWp \rj\rj+X \to \Pep\Pne\mu^+\nu_\mu \rj\rj+X$}.  This
process leads to a distinct signature of same-sign high-$\pT$ leptons,
missing energy and jets. Since no gluon-initiated processes
contribute to this final state at LO,
it has a comparably low SM cross section and thus is a good candidate
to search for physics beyond the SM.  New-physics signals involving
same-sign leptons originate for instance in $R$-parity-violating SUSY
models~\cite{Dreiner:2006sv}, in di-quark production with decay of the
di-quark to a pair of top quarks~\cite{Han:2009ya}, or from the
production of doubly-charged Higgs bosons~\cite{Maalampi:2002vx}.
Moreover, it constitutes a background to double parton scattering
\cite{Kulesza:1999zh,Maina:2009sj,Gaunt:2010pi}.

Since LO cross sections carry a large uncertainty, the calculation of the 
NLO corrections in the strong coupling is needed to obtain a reliable 
prediction. For the QCD-mediated contributions to
\mbox{$\Pp\Pp\to \PWp\PWp \rj\rj+X\to \Pep\Pne\mu^+\nu_\mu \rj\rj+X$}, 
and later also for 
\mbox{$\Pp\Pp\to \PWp\PWm\rj\rj+X\to \Pep\Pne\bar{\nu}_\mu \mu^-\rj\rj+X$}, 
NLO results were
presented in double-pole approximation, i.e.\ including only diagrams
with two resonant $\PW$ bosons, but leptonic $\PW$~decays with full spin
correlations~\cite{Melia:2010bm,Melia:2011dw,Greiner:2012im}. 
For $\PW^+\PW^+$ the computation~\cite{Melia:2010bm} has been subsequently 
implemented~\cite{Melia:2011gk}
into the \textsc{POWHEG BOX}~\cite{Nason:2004rx,Frixione:2007vw}.
In a series of NLO calculations for vector-boson scattering processes
in VBF~\cite{Jager:2006zc,Jager:2006cp,Bozzi:2007ur,Jager:2009xx}, NLO
results for the EW production mode were given for the complete process
\mbox{$\Pp\Pp\to \PWp\PWp \rj\rj+X\to \Pep\Pne\mu^+\nu_\mu\rj\rj+X$} in
\citere{Jager:2009xx} including the full set of $t$- and $u$-channel
diagrams (also those without resonant W~bosons) while neglecting
$s$-channel diagrams and interferences between $t$- and $u$-channel
contributions.  Also this computation has recently been combined with
a parton shower~\cite{Jager:2011ms} using the \textsc{POWHEG BOX}.

In this work we present an independent calculation of the NLO QCD
corrections to EW $\PWp\PWp\rj\rj$ production including leptonic
$\PW$-boson decays and non-resonant diagrams. This constitutes the
first independent check of the calculation in \citere{Jager:2009xx}.
Moreover, we investigate the size of the $s$-channel and interference
contributions at LO.

The paper is organized as follows: In \refse{method} we discuss
technical aspects of the calculation, namely the organization of
Feynman diagrams
into building blocks (\refse{chaptstructure}) and the
evaluation of the NLO corrections (\refse{chaptnlo}). 
\refse{comparison} covers numerical checks and comparisons with
previously published results.  Finally, in \refse{chaptresults}
we present numerical predictions for the LHC at $14\TeV$, while
\refse{chaptsummary} contains our conclusions.

\section{Elements of calculation}\label{method}

We have developed a general framework for the calculation of QCD
corrections to vector-boson scattering reactions at hadron colliders
\cite{Lucia}, i.e.\ for EW processes of the form \mbox{$\Pp\Pp\to 4l\rj\rj+X$}
with 4 arbitrary (charged and/or neutral) leptons. In this section we
sketch the ingredients and main features of the method.

Because the LHC experiments are conducted at TeV energies,
fermion-mass effects are strongly suppressed and have been neglected.
At the same time, only the two lighter generations of quarks (u, d, c
and s) and leptons have been taken into account. In
\citere{Ciccolini:2007ec}, the contribution of external b quarks to
Higgs production via VBF has been found to be at the level of 2\% if
VBF cuts are applied (4\% without VBF cuts). For the processes
involved in our calculation, these contributions can be expected to be
of similar size, if not even smaller: e.g.\ in the $\PWp\PWp$ channel
discussed in this paper, external bottom \mbox{(anti-)quarks} would
show up only accompanied by non-diagonal CKM matrix elements and thus
be entirely negligible. Further, it can be demonstrated
\cite{Bozzi:2007ur} that the CKM matrix can be approximated by a unit
matrix provided the interferences between different kinematic channels
as well as the $s$-channel contribution are negligible, which is
verified in \refse{chaptresults}.

\subsection{Structure of the diagrams and building blocks}\label{chaptstructure}

For the calculation of the LO and NLO matrix elements of the processes
\mbox{$\rp \rp \to 4l\rj\rj+X$} we adopt a similar strategy as in
\citeres{Jager:2006zc,Jager:2006cp,Bozzi:2007ur,Jager:2009xx}. In
order to deal with the large number of diagrams%
\footnote{For \mbox{$\Pu\Pc\to \Pd\Ps\Pep\Pne\mu^+\nu_\mu$} there are
93 diagrams at LO. At NLO, 430 loop and 346 counterterm diagrams
contribute, while the basic partonic real-emission process
\mbox{$\Pu\Pc\to \Pd\Ps\Pep\Pne\mu^+\nu_\mu\Pg$} contains 452
diagrams.},
we introduce generic building blocks from which the matrix elements
can be constructed. The details of our approach differ, however, in
many aspects from
\citeres{Jager:2006zc,Jager:2006cp,Bozzi:2007ur,Jager:2009xx}.

\begin{figure}
\centering
\subfloat[Type A]{\centering\label{typeA}\includegraphics[bb = -5 572 595 772, height=4cm]{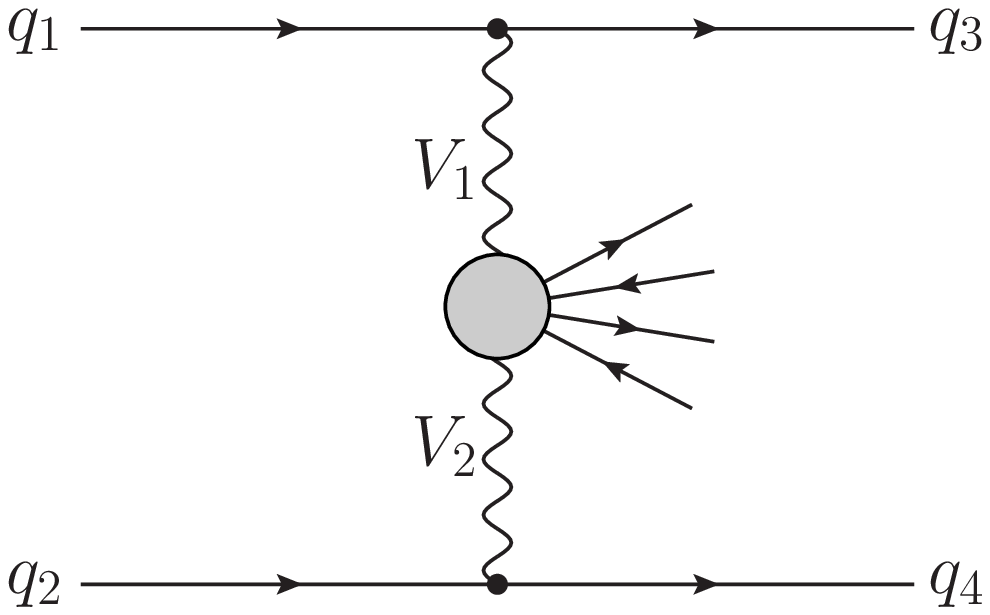}}\\
\subfloat[Type B]{\label{typeB}\includegraphics[bb = 145 572 445 772, height=4cm]{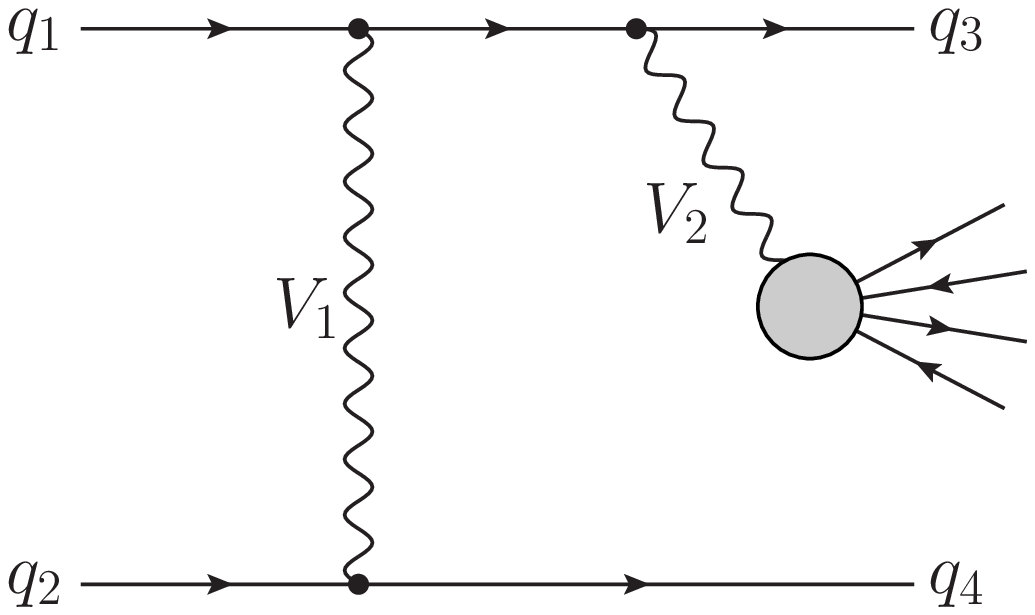}}\qquad\quad
\subfloat[Type C]{\label{typeC}\includegraphics[bb = 145 572 445 772, height=4cm]{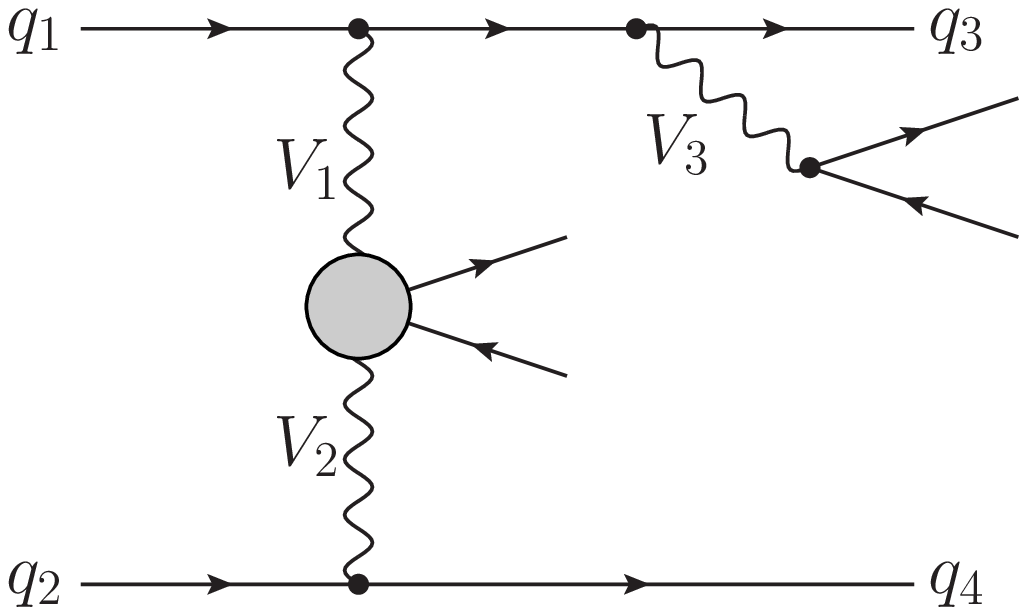}}\\
\subfloat[Type D]{\label{typeD}\includegraphics[bb = 145 572 445 772, height=4cm]{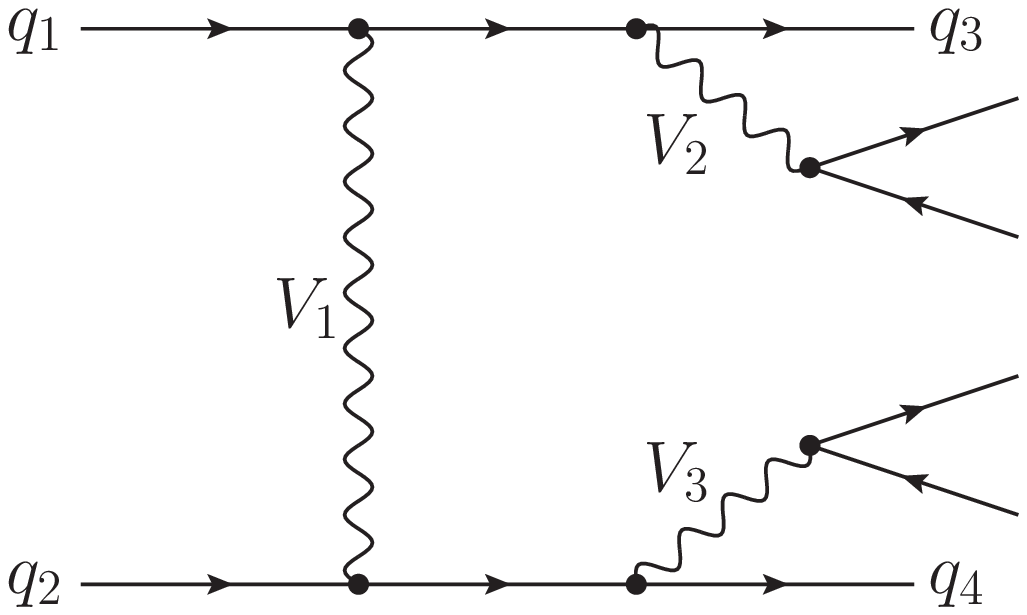}\qquad\quad\includegraphics[bb = 145 572 445 772, height=4cm]{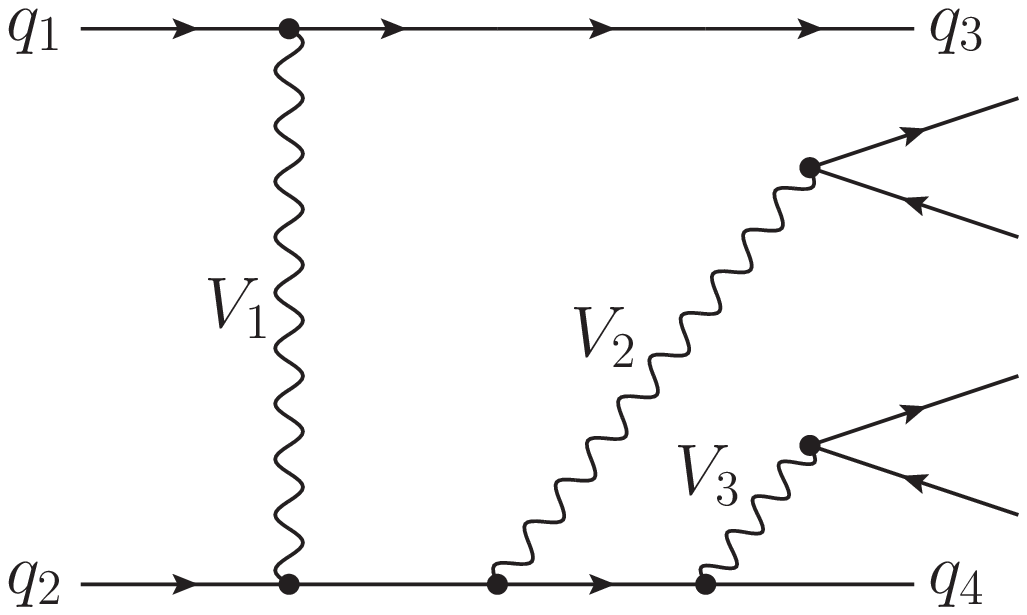}}
\centering
\caption{Generic types of {\it t}-channel topologies}
\label{types}
\end{figure}

For the class of processes \mbox{$\rp \rp \to 4l\rj\rj+X$}, the
Feynman diagrams can be divided into four generic categories, taking
advantage of the fact that the EW and QCD parts of the diagrams are
largely independent of one another. \reffi{types} demonstrates four
generic types into which all {\it t}-channel diagrams involved in our
calculation can be categorized.  

Type A (\reffi{typeA}) represents the genuine VBF
diagram, with two vector bosons radiated off the quark lines
fusing in the centre to produce four leptons in the final state.

Type B (\reffi{typeB}) contains two quark lines connected with a
vector boson and another vector boson radiated off either of the two
quark lines which decays via EW interactions into four
final-state leptons.
For combinatorial reasons, 4 topologies of this type exist.

Type C (\reffi{typeC}) has a vector boson radiated off either one of
the two quark lines which decays into two leptons, and two more vector
bosons fusing in the central region to produce a second pair of
leptons. Again, 4 topologies of this type exist.

Finally, type D (\reffi{typeD}) sees one vector boson connecting the
quark lines, and two more radiated either one off each quark line
(diagram on the left) or both off the same quark line (diagram on the
right) and subsequently decaying into two lepton pairs. In total,
there are 10 different topologies of this type which are grouped
together as they involve the same EW building blocks.

Each generic diagram involves two QCD parts---the two quark lines with
attached vector bosons---and one or two EW parts, namely the vector-boson scattering
block and/or vector-boson decays into leptons. Owing
to charge conservation, not all generic topologies give rise to
Feynman diagrams once particular insertions for the external
\mbox{(anti-)quarks}, the final-state leptons, and, correspondingly, the
intermediate EW bosons are fixed. Thus, for instance, type B is
completely absent if the final-state lepton charges add up to $\pm2$.

Evidently changes to either QCD or EW parts that do not alter
the momenta of the internal vector bosons have no effect on the rest of
the diagram. For instance, application of crossing symmetry to the
upper quark line does not influence the lower quark line and all
leptonic parts.  Similar arguments hold for adding a gluon loop to
either of the quark lines, which essentially amounts to calculating
virtual NLO QCD corrections to the entire diagram. Since the leptonic
sector of the diagram in itself can be quite complicated, it is
advantageous to calculate these blocks only once and reuse them with
different QCD parts.

The calculation is performed in the 't Hooft--Feynman gauge.  Diagrams
with would-be Goldstone bosons connecting EW and QCD parts do not
contribute because their couplings to massless fermions vanish.%
\footnote{Note that would-be Goldstone bosons show up inside the EW
vector-boson scattering building block of \reffi{typeA},
namely in all processes involving a $\PWp\PWm\PZ/\gamma\PZ/\gamma$ vertex.}
Factorization of parts of the diagrams can be achieved by inserting
the polarization sums for massive vector bosons,
\beq\label{polsum}
g^{\mu\nu}=-\sum_{i=\{+,-,0\}}\varepsilon^\mu_i(k)\varepsilon^{*\nu}_i(k)+\f{k^\mu k^\nu}{k^2}\;,
\eeq
for the numerators of the gauge-boson propagators coupled to a quark
line, effectively thus cutting the diagram into blocks that can be
evaluated on their own. The polarization vectors
$\varepsilon^\mu_i(k)$ and $\varepsilon^{*\nu}_i(k)$ for off-shell
particles are obtained by replacing the vector-boson mass with
$\sqrt{k^2}$ in the definition of the longitudinal polarization
vector, i.e.\ 
\beq
\varepsilon_0^\mu(k)=\f{k^0}{\sqrt{k^2}}\left(\f{|\mathbf k|}{k^0},\cos\phi\sin\theta,\sin\phi\sin\theta,\cos\theta\right).
\eeq
The polarization vectors $\varepsilon^{\mu}_{\pm}(k)$ do not depend on
the mass and thus remain unchanged.  Introducing
\beq
\varepsilon_m^\mu(k)=\f{k^\mu}{\sqrt{k^2}},\qquad
\varepsilon_m^{*\mu}(k)=-\f{k^\mu}{\sqrt{k^2}},
\eeq
to compactify the notation,
the polarization sum (\ref{polsum}) can be rewritten as
\beq\label{polsumshort}
g^{\mu\nu}=-\sum_{i=\{+,-,0,m\}}\varepsilon^\mu_i(k)\varepsilon^{*\nu}_i(k).
\eeq
Because of gauge invariance the contractions of $k^\mu$ with some of
the building blocks vanish.  After checking numerically that these
terms do not contribute, we have omitted them in those type D diagrams
(\reffi{typeD}) in which two outgoing vector bosons couple to the same
quark line since their evaluation consumes most CPU time (in
comparison to the remaining topologies), particularly at NLO.

\begin{figure}
\begin{adjustwidth}{-3em}{-2em}
\centering
\includegraphics[bb = 145 572 595 772, height=4cm]{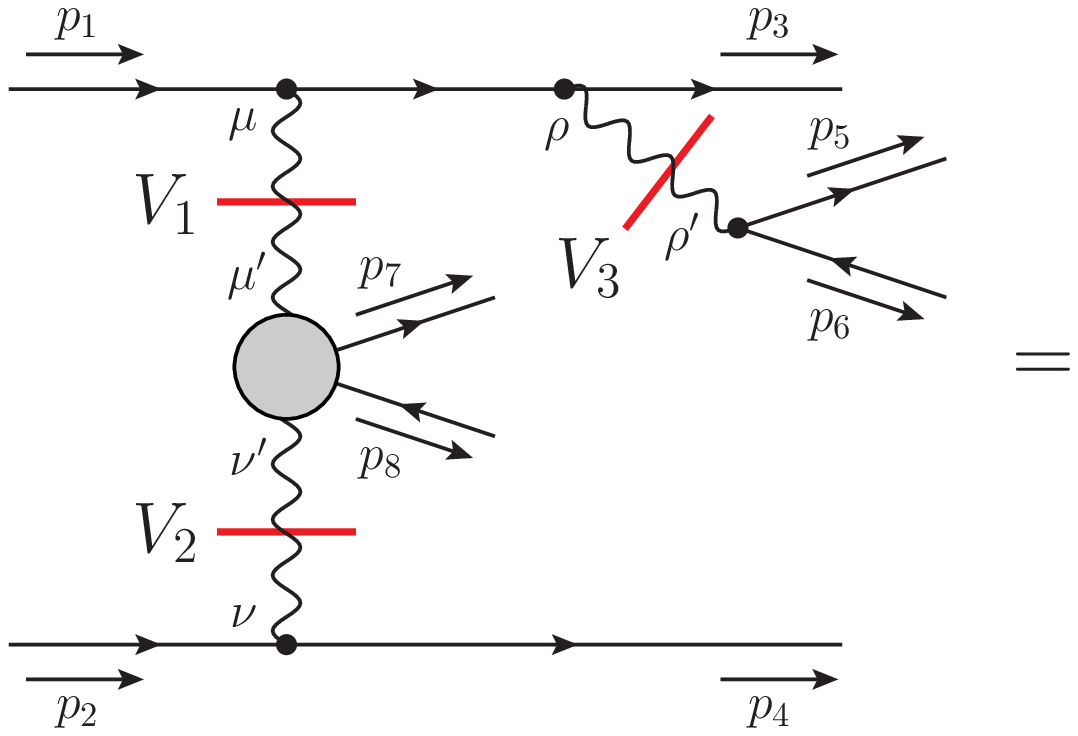}\quad
\includegraphics[bb = 195 572 545 772, height=4cm]{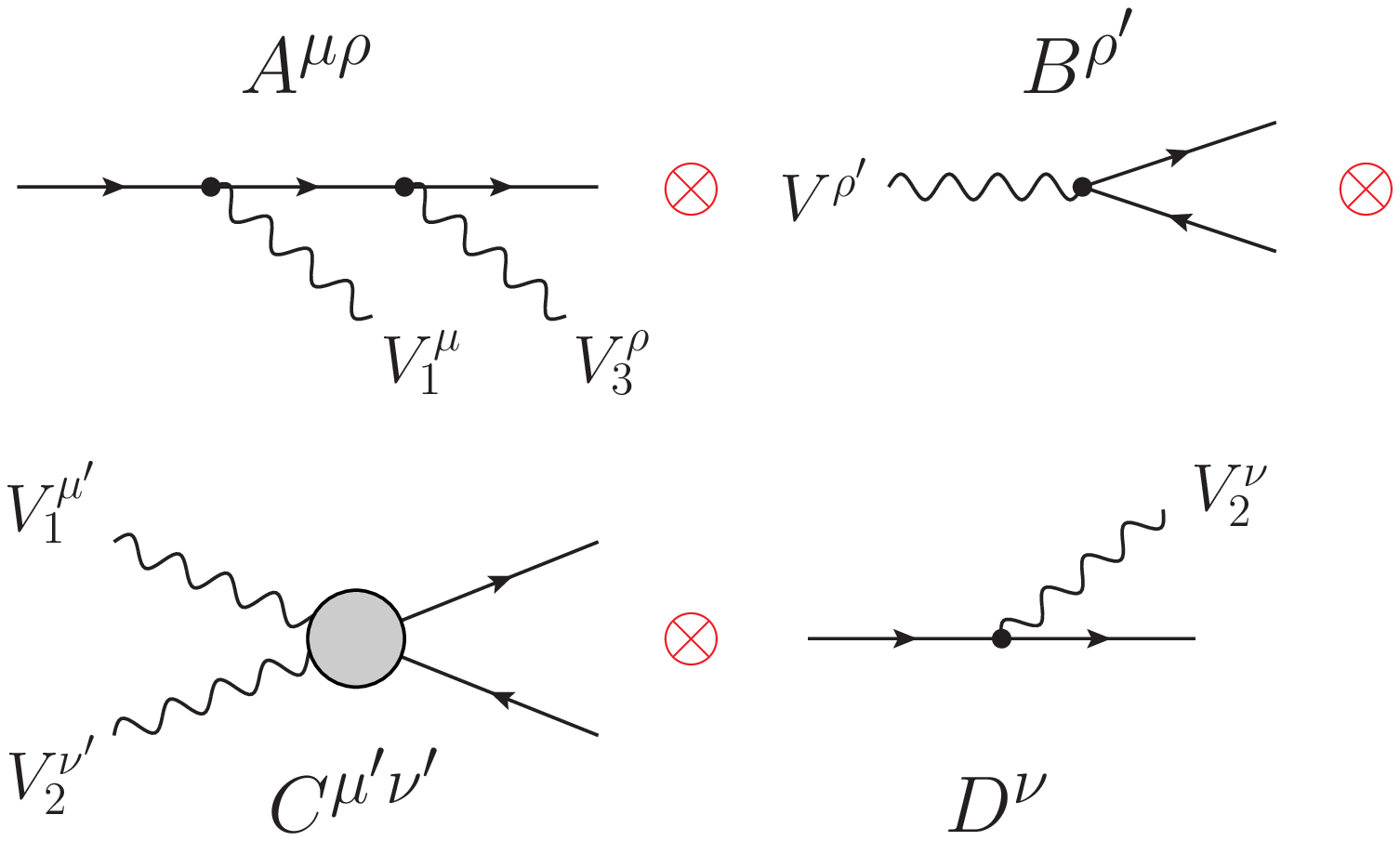}\\[8ex]
\centering
\end{adjustwidth}
\caption{ Example of a diagram split into four building blocks by applying the polarization sums
  (\ref{polsum}) to cut three intermediate vector bosons.}
\label{polsum2}
\end{figure}

Implementing the block structure by cutting all internal vector bosons
that couple to the quark lines in the diagrams in \reffi{types} allows
us to not only save CPU time by evaluating each required block only
once, but also to keep the number of required blocks relatively small
by reusing them in multiple instances throughout all diagrams and even
partonic processes. The diagrams we need to consider in our calculation
contain up to three vector bosons that are being radiated off the
quark lines, and the polarization sum has to be applied once to each
of their propagators. \reffi{polsum2} illustrates how a diagram
featuring three vector bosons can be split into four building blocks.
Each splitting represents an insertion of one polarization sum.  The
resulting amplitude reads:
\begin{align}\label{eq:ampsplit}
-\sum_{i,j,k={\pm,0,m}}& [A^{\mu\rho}\varepsilon^*_{1,i,\mu}(k_1)\varepsilon^*_{3,k,\rho}(k_3)]\;[B^{\rho'}\varepsilon_{3,k,\rho'}(k_3)]\;[C^{\mu'\nu'}\varepsilon_{1,i,\mu'}(k_1)
\varepsilon_{2,j,\nu'}(k_2)]\;[D^\nu\varepsilon^*_{2,j,\nu}(k_2)]\nn\\
&\times \frac{1}{k_1^2-M_{V_1}^2}  \frac{1}{k_2^2-M_{V_2}^2}\frac{1}{k_3^2-M_{V_3}^2},
\end{align}
where \mbox{$1/{(k_i^2-M_{V_i}^2)}$} are the denominator parts of the gauge-boson
propagators, and  \mbox{$k_1=p_1-p_3-p_5-p_6$}, \mbox{$k_2=p_2-p_4$}, and \mbox{$k_3=p_5+p_6$},
according to the notation introduced in \reffi{polsum2}.

\begin{figure}
\centering
\subfloat[\hspace*{-1.8em}]{\label{Vll}\includegraphics[bb = 215 692 380 772, height=2.5cm]{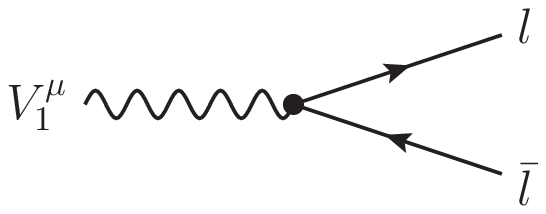}}\qquad\qquad
\subfloat[\hspace*{-1.8em}]{\label{Vllll}\includegraphics[bb = 215 692 380 772, height=2.5cm]{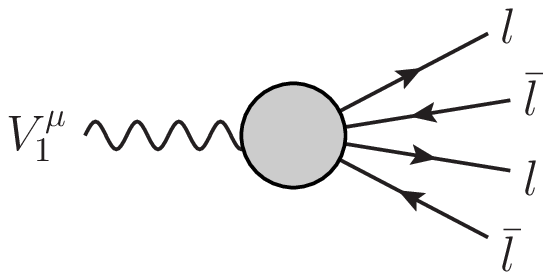}}\\
\subfloat[\hspace*{-1.8em}]{\label{VVll}\includegraphics[bb = 215 692 380 772, height=2.5cm]{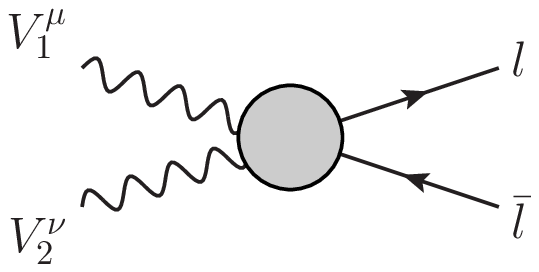}}\qquad\qquad
\subfloat[\hspace*{-1.8em}]{\label{VVllll}\includegraphics[bb = 215 692 380 772, height=2.5cm]{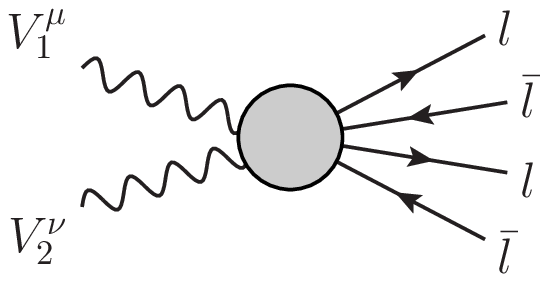}}\\
\centering
\caption{Building blocks involving leptons.}
\label{lepttensors}
\end{figure}

Building blocks involving leptons (shown in \reffi{lepttensors})
typically involve more than one Feynman diagram, with the exception of
the block corresponding to the EW current (\reffi{Vll}).
Diagrams of type B (\reffi{typeB}) contain building blocks with
one vector boson in the initial state and four leptons in the final
state (\reffi{Vllll}).  Building blocks with two external vector
bosons (\reffis{VVll}{VVllll}) are represented by a
\mbox{$4\times 4$} array, each element corresponding to one term of the
complete polarization sum constructed by cutting the two vector bosons.

\begin{figure}
\centering
\subfloat[\hspace*{1.8em}]{\label{qqV}\includegraphics[bb = 220 692 400 772, height=2.5cm]{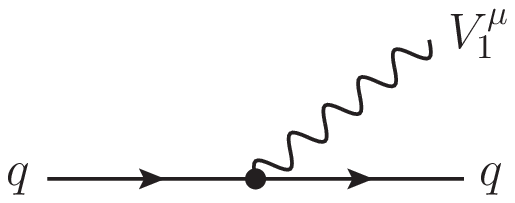}}\qquad\quad
\subfloat[\hspace*{1.8em}]{\label{qqVV}\includegraphics[bb = 220 692 400 772, height=2.5cm]{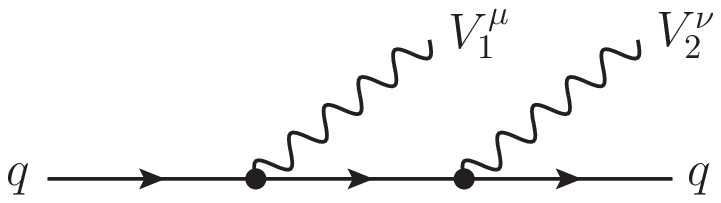}}\\
\subfloat[\hspace*{-.2em}]{\label{qqVVV}\includegraphics[bb = 220 692 400 772, height=2.5cm]{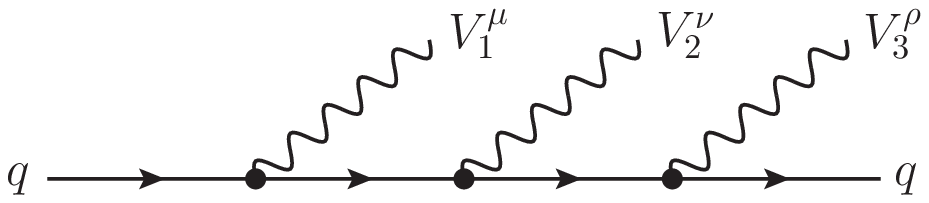}}
\caption{LO building blocks involving quark lines.}
\label{LOblocks}
\end{figure}

At LO, the QCD building blocks are formed by one diagram each, as
shown in \reffi{LOblocks}.  Blocks involving two vector bosons
entering diagrams of type B, C and D are represented by a
\mbox{$4\times 4$} array (\reffi{qqVV}).  Building blocks with three
outgoing vector bosons (\reffi{qqVVV}) appearing in type D are
represented by \mbox{$4\times 4\times 4$} arrays.

\begin{table}[!htb]
\centering
\begin{tabular}[]{ccccc}
$\rW^+\rW^+$: & $\ru\rc\to  \rd\rs  \re^+\nu_{\rm e}\mu^+\nu_\mu$ &\hspace*{3em}& $\rW^+\rW^-$: & $\ru\rc \to  \ru\rc  \re^+\nu_{\rm e}\bar\nu_\mu\mu^-$ \\
  &  &&  & $\rd\rs\to  \rd\rs  \re^+\nu_{\rm e}\bar\nu_\mu\mu^-$ \\
$\rW^-\rW^-$: & $\rd\rs\to  \ru\rc   \bar\nu_{\rm e}\re^-\bar\nu_\mu\mu^-$ &&  & $\ru\rs\to  \ru\rs  \re^+\nu_{\rm e}\bar\nu_\mu\mu^-$ \\
  &  &&  & $\ru\rs\to  \rd\rc  \re^+\nu_{\rm e}\bar\nu_\mu\mu^-$ \\
$\rW^+\rZ$: & $\ru\rc\to  \rd\rc   \re^+\nu_{\rm e}\mu^+\mu^-$ &&  & \\
  & $\ru\rs\to  \rd\rs   \re^+\nu_{\rm e}\mu^+\mu^-$ && $\rZ\rZ$: & $\ru\rc\to  \ru\rc  \re^+\re^-\mu^+\mu^-$\\
  &  &&  & $\rd\rs\to  \rd\rs  \re^+\re^-\mu^+\mu^-$ \\
$\rW^-\rZ$: & $\rd\rc\to  \ru\rc   \bar\nu_{\rm e}\re^-\mu^+\mu^-$ &&  & $\ru\rs\to  \ru\rs  \re^+\re^-\mu^+\mu^-$ \\
  & $\rd\rs\to  \ru\rs   \bar\nu_{\rm e}\re^-\mu^+\mu^-$ &&  & $\ru\rs\to  \rd\rc  \re^+\re^-\mu^+\mu^-$\\
\end{tabular}
\vspace*{2ex}
\caption{List of generic $t$-channel matrix elements corresponding to the 
intermediate weak bosons produced in the fusion diagrams. 
Final states with less charged leptons are obtained by modifying the
leptonic building blocks only, i.e.\ by replacing $\mu^+\mu^-$ by
$\bar{\nu}_\mu\nu_\mu$ or $\re^+\re^-\mu^+\mu^-$ by
$\bar{\nu}_{\rm e}\nu_{\rm e}\bar{\nu}_\mu\nu_\mu$. Moreover, $\rW^+\rW^-$ and $\rZ\rZ$
in general mix if same-flavour leptonic final states like
$\re^+\nu_\re\bar\nu_\re\re^-$ are considered. 
All partonic processes contributing to the respective hadronic
cross sections can be obtained from these generic matrix elements.}
\label{listall}
\end{table}

All partonic processes contributing to a specific process
\mbox{$\Pp\Pp\to \PV\PV\rj\rj+X\to 4l\rj\rj+X$} 
can be constructed from up to 8 generic $t$-channel matrix elements 
listed in \refta{listall} 
(e.g.\ \mbox{$\ru\rc\to  \rd\rs \re^+\nu_{\rm e}\mu^+\nu_\mu$} for $\PWp\PWp$) 
by applying crossing symmetry to reverse the flow of either one or both quark 
currents (e.g.\ \mbox{$\ru\bar\rs\to  \rd\bar\rc \re^+\nu_{\rm e}\mu^+\nu_\mu$})
and to construct the $s$-channel diagrams 
(e.g.\ \mbox{$\bar \rs\rc\to \rd\bar \ru\re^+\nu_{\rm e}\mu^+\nu_\mu$}), or by
exchanging the outgoing lines to obtain $u$-channel diagrams 
(e.g.\ \mbox{$\ru\rc\to  \rs\rd \re^+\nu_{\rm e}\mu^+\nu_\mu$}).  
As the order of the outgoing partons is obviously arbitrary, the
distinction between $t$- and $u$-channel diagrams only makes sense if
both types contribute to the same partonic process 
(e.g.\ \mbox{$\ru\ru\to  \rd\rd \re^+\nu_{\rm e}\mu^+\nu_\mu$}), which
is the case if all \mbox{(anti-)quarks} belong to the same generation.

With the CKM matrix approximated by a unit matrix, partonic processes 
which result from one another by interchanging all first-generation
\mbox{(anti-)quarks} with their second-generation counterparts and
vice versa, are described by the same matrix elements.  For instance,
the partonic processes 
\mbox{$\ru\ru\to  \rd\rd\re^+\nu_{\rm e}\mu^+\nu_\mu$} and
\mbox{$\rc\rc\to  \rs\rs\re^+\nu_{\rm e}\mu^+\nu_\mu$}
only differ in the parton distribution functions,
and the matrix element can be recycled.
Analogously, partonic processes 
involving two different generations of quarks---for example
\mbox{$\bar\rs\rc\to  \bar \ru\rd\re^+\nu_{\rm e}\mu^+\nu_\mu$} and
\mbox{$\bar\rd\ru\to  \bar \rc \rs\re^+\nu_{\rm e}\mu^+\nu_\mu$}---are pairwise
formed by identical matrix elements.

In our calculation, the formulae for combining the blocks have been
implemented as \textsc{Fortran} subroutines.  The expressions for the
individual building blocks are obtained by means of the
\textsc{FormCalc} 6 package~\cite{Hahn:2009bf}, which also introduces
abbreviations for the fermion chains and thus helps to speed up the
code significantly. They are evaluated using the Weyl--van der Waerden
(WvdW) helicity formalism~\cite{Dittmaier:1998nn} allowing us to
express all subamplitudes involved in the polarization sums in terms
of universal WvdW spinors and compute them numerically. The
$\textsc{FormCalc}$ code is modified to transform the amplitudes to
the form \refeq{eq:ampsplit}, further abbreviations of spinor products
are introduced, and each building block is exported into a
$\textsc{Fortran}$ module which takes the momenta and helicities for
the particles in the building block as input.  The \textsc{Fortran}
code for each process is contained in a single function that can be
called from within a Monte Carlo program and returns an array of full
squared amplitudes for each partonic process, 
including all relevant colour and averaging factors.

\subsection{Calculation of NLO cross sections and matrix elements}\label{chaptnlo}

Cross sections involving two initial-state hadrons at a fixed
perturbative order are given as a convolution of the parton
distribution functions (PDFs) and the partonic cross sections $\hat
\sigma_{ab}$, summed over all incoming partons resulting in
contributions to the considered hadronic process.

At LO, the cross section is defined by
\beqa\label{LOcrosssection}
\sigma_{\Pp\Pp}^{\mathrm{LO}} &=&\sum_{a,b}\int_0^1\rd x_1\int_0^1\rd x_2\,
f^{\mathrm {LO}}_a(x_1,\muF)f_b^{\mathrm{LO}}(x_2,\muF)
\int_m\rd\Phi_m\, \rd\hat\sigma^{\mathrm B}_{ab}(x_1p_1,x_2p_2),
\eeqa
where \mbox{$f_{a/b}(x_{1/2},\muF)$} are the
PDFs that give the probability to find parton $a/b$ with a momentum
fraction $x_{1/2}$ in the respective proton,
\mbox{$\rd\hat\sigma^{\mathrm B}_{ab}(x_1p_1,x_2p_2)$}
is the differential partonic Born cross section which is integrated
over the $m$-parton phase-space $\Phi_m$.
While the sum over the incoming partons $a$ and $b$ is explicitly
stated, summation over all outgoing parton configurations giving rise
to non-vanishing partonic contributions to the hadronic process
discussed is implicitly assumed.

At NLO QCD, virtual and real corrections contribute to the cross section, which
separately contain soft and collinear divergences.
However,
these infrared (IR) divergences cancel in the NLO cross section, if
IR-safe jet observables are considered and the PDFs are renormalized appropriately.
For mediating this cancellation, the Catani--Seymour dipole-subtraction technique
for massless particles~\cite{Catani:1996vz} is applied. This procedure allows us to
express the NLO cross section as a sum over individually finite phase-space integrals,
{\allowdisplaybreaks
\begin{align}\label{cross section dipole}
\raggedright\nn
\sigma_{\Pp\Pp}^{\mathrm{NLO}} =&\sum_{a,b}\int_0^1\rd x_1\int_0^1\rd
x_2\,
f^{\mathrm {NLO}}_a(x_1,\muF)f_b^{\mathrm {NLO}}(x_2,\muF)\nn\\
&{}\times\Biggl\{\int_m\rd\Phi_m\bigl[\rd\hat\sigma^{\mathrm B}_{ab}(x_1p_1,x_2p_2)+
    \rd\hat\sigma^{\mathrm V}_{ab}(x_1p_1,x_2p_2)+\mathbf{I}\otimes\rd\hat\sigma^{\mathrm B}_{ab}(x_1p_1,x_2p_2)\bigr]\nn\\
&{}+\int_0^1\rd z_1\int_m\rd\Phi_m\;\Bigl(\mathbf{K}_{aa^\prime}(z_1)+\mathbf{P}_{aa^\prime}(z_1)\Bigr)\otimes\rd\hat\sigma^{\mathrm B}_{a^\prime b}(z_1 x_1p_1,x_2p_2)\nn\\
&{}+\int_0^1\rd z_2\int_m\rd\Phi_m\;\Bigl(\mathbf{K}_{bb^\prime}(z_2)+\mathbf{P}_{bb^\prime}(z_2)\Bigr)\otimes\rd\hat\sigma^{\mathrm B}_{ab^\prime}(x_1p_1,z_2 x_2p_2)\nn\\
&+{}\int_{m+1}\rd\Phi_{m+1}\left(\rd\hat\sigma_{ab}^{\mathrm
      R}(x_1p_1,x_2p_2)-\sum_{\mathrm{dipoles}}(\rd V_{\mathrm{dipole}}\otimes\rd\hat\sigma^{\mathrm
      B})_{ab}(x_1p_1,x_2p_2)\right)\Biggr\},\nn\\
\end{align}
}%
with the conventions of \refeq{LOcrosssection}. The differential
partonic contributions
$\rd\hat\sigma^{\mathrm V}_{ab}$ and $\rd\hat\sigma^{\mathrm R}_{ab}$
correspond to the virtual and real corrections, respectively.
The process-independent operators $\mathbf{I}$, $\mathbf{K}$, and $\mathbf{P}$
are defined in~\citere{Catani:1996vz}, and $\otimes$ symbolizes the colour correlations between
Born matrix elements and these operators (spin correlations do not appear in the
given process class as no external gluons are involved at Born level).
In the processes \mbox{$\Pp\Pp\to \PV\PV\rj\rj+X$}, 
the Born and virtual cross
sections are built from the partonic initial states $qq,\bar{q}q,q\bar{q},\bar{q}\bar{q}$,
while in the real cross section additionally $qg,gq,\bar{q}g,g\bar{q}$
contribute.

The process-dependent ingredients which are necessary for calculating the NLO cross sections
of the processes \mbox{$\Pp\Pp\to \PV\PV\rj\rj+X$} are thus:
\begin{enumerate}
\item 
Born-level level matrix elements $\mathcal{M}^{\rm B}$ needed for $\mathrm{d}\hat\sigma^{\rm B}$,
evaluated in four dimensions. Their construction is outlined in the previous section.
\item 
One-loop virtual matrix elements $\mathcal{M}^{\rm V}$ needed
for $\mathrm{d}\hat\sigma^{\rm V}$, with renormalized ultraviolet
divergences and IR divergences regularized using dimensional
regularization, evaluated in $D$ dimensions. Once the QCD and EW
sections of the diagrams are separated, the transition from LO to
virtual corrections can be performed by adding a gluon loop to
either of the two quark lines. Continuing with the example in
\reffi{polsum2}, one of the building blocks $A^{\mu\rho}$ or
$D^\nu$ is replaced by $A_{\mathrm{V}}^{\mu\rho}$ or $D_{\rm V}^\nu$
(shown in \reffi{virtblocks}) respectively, while the other one
remains unchanged.  The ultraviolet divergences are renormalized by
adding the corresponding counterterms.  Since neither of these changes
has any influence on the overall kinematics of the diagram, the
leptonic blocks $B^{\rho'}$ and $C^{\mu'\nu'}$ stay the same as in the
LO.
\item 
Real-radiation matrix elements $\mathcal{M}^{\rm R}$ needed for
$\mathrm{d}\hat\sigma^{\rm R}$, evaluated in four dimensions. They
can be created in a similar manner as in the LO case.
In the example from \reffi{polsum2}, this amounts to
attaching an outgoing gluon in every possible way in the building
blocks $A^{\mu\rho}$ or $D^\nu$ and shifting the momenta $k_1$,
$k_2$ and $k_3$ of the intermediate vector bosons accordingly.
The diagrams with an initial-state gluon can then be obtained via
crossing symmetry.

\begin{figure}
\centering
\includegraphics[bb = 920 450 0 600,scale=0.55]{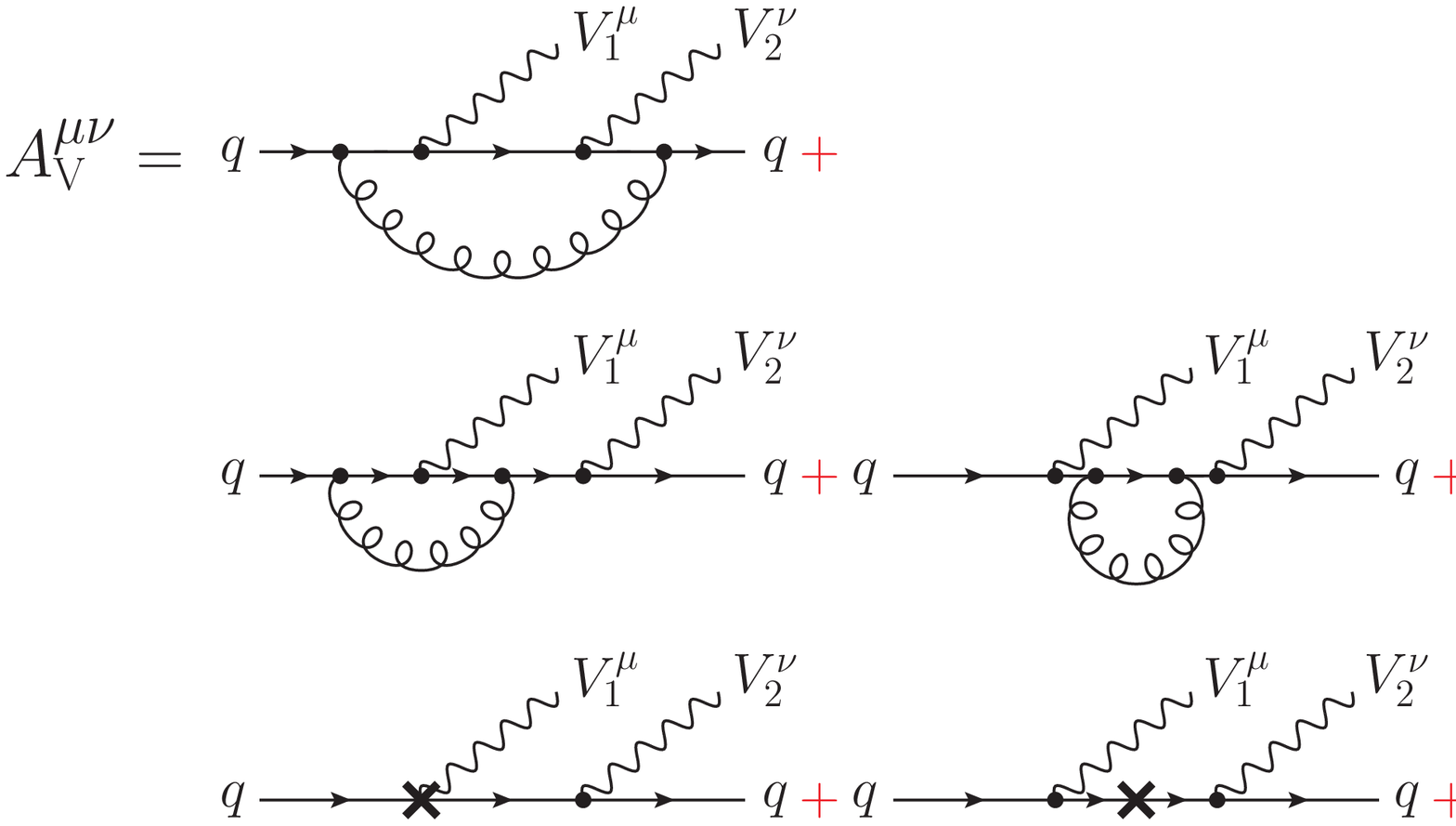}\\
\includegraphics[bb = 920 870 0 990,scale=0.55]{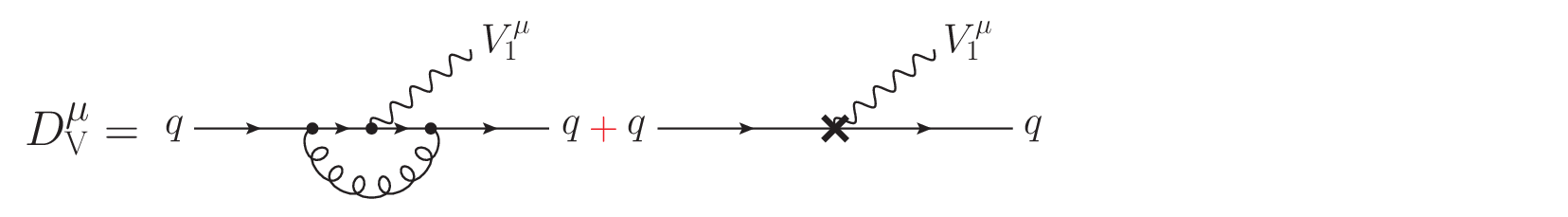}\\[20ex]
\centering
\caption{Virtual contributions $A^{\mu\nu}_{\rm V}$ and $D^\mu_{\rm V}$ corresponding to building blocks $A^{\mu\nu}$ and $D^\mu$ in Figure \ref{polsum2}.}
\label{virtblocks}
\end{figure}

\item 
A set of colour projected Born-level matrix elements required
to construct
\linebreak\mbox{$\rd V_{\mathrm{dipole}}\otimes\rd\hat\sigma^{\mathrm B}$} and
\mbox{$(\mathbf{K}+\mathbf{P})\otimes\rd\hat\sigma^{\mathrm B}$},
evaluated in four dimensions.  In the VBF approximation of the matrix
elements (no $s$-channel diagrams and no interferences between $t$-
and $u$-channel diagrams), the colour correlations turn out to be
trivial and give rise to the same constant factor $-C_{\mathrm{F}}$
for all colour-correlated Born matrix elements.
\end{enumerate}

Analytical expressions for the virtual amplitudes have been generated
in the same way as for the Born amplitudes.  All divergences of the
loop diagrams appear in tensor integrals, which are given by two,
three, four and five-point functions.  In our calculation, the tensor
reduction is performed numerically
in \textsc{Fortran} by means of the \textsc{COLLIER}
library~\cite{Denner:2002ii,Denner:2005nn,Denner:1991qq,Beenakker:1988jr,Denner:2010tr,Coli6}
which is based on the tensor reduction scheme developed by Denner and
Dittmaier~\cite{Denner:2002ii,Denner:2005nn} and supports both mass and dimensional
regularization scheme.

The resonant \PW\ bosons require a proper inclusion of the finite
vector-boson widths in the propagators. We use the complex-mass
scheme, which was introduced in \citere{Denner:1999gp} for LO
calculations and generalized to the one-loop level in
\citere{Denner:2005fg}. In this approach, the W- and Z-boson masses as
well as the Higgs-boson mass are consistently considered complex
quantities, defined as the locations of the propagator poles in the
complex plane.  This leads to complex couplings and, in particular, a
complex weak mixing angle.  The scheme fully respects all relations
that follow from gauge invariance. A brief description of the
complex-mass scheme can also be found in \citere{Denner:2006ic}.

\section{Checks and comparison with existing results}\label{comparison}

In order to verify the correctness of the calculation, comparisons with
available results and tools have been performed at each step.  The
matrix elements for each partonic process have been verified for a set of
phase-space points. In the zero-width limit, we have compared the Born
and real-correction amplitudes for individual partonic processes 
with \textsc{MadGraph~4}~\cite{Alwall:2007st} at single precision accuracy
as well as with stand-alone $\textsc{FormCalc}$ 6~\cite{Hahn:2009bf}
at double-precision accuracy and found full agreement within the
numerical accuracy of the calculation.  For virtual corrections, the
matrix elements have been generated with tensor reduction performed in
both \textsc{COLLIER} and \textsc{LoopTools}, using mass
regularization for IR divergences. The two results agreed at the
$10^{-8}$ level. Cancellation of the UV divergences has been tested
numerically by varying the value of the UV regulator
$\varepsilon_{\mathrm{UV}}$ from $10^{-5}$ to $10^{5}$; the resulting
amplitude stays unchanged up to the level of $10^{-11}$. Born and
real matrix elements in the complex-mass scheme have been checked
against \textsc{OpenLoops}~\cite{Cascioli:2011va} and found to be in
full agreement within double-precision accuracy, both for full
amplitudes (without approximations) as well as for the amplitudes
in the so-called VBF approximation (neglecting $s$-channel diagrams
and interferences between channels); in the \textsc{OpenLoops}
framework, the VBF approximation was imposed by selecting the relevant
parts of the squared matrix elements according to their particular
colour structure.

Furthermore, the pole structure of the virtual corrections is given by
the following formula~\cite{Jager:2006zc} derived from the
$\mathbf{I}$ operator\cite{Catani:1996vz}:
\beq\label{virtcheck}
\mathcal{M}^{\rm{V}}_{\rm{block}}=\mathcal{M}^{\rm  B}_{\rm{block}}\f{\alpha_\rs}{4\pi}C_{\rm F}
\left(\f{1}{Q^2}\right)^{\varepsilon_{\rm{IR}}}
\left[-\f{2}{\varepsilon_{\rm{IR}}^2}-\f{3}{\varepsilon_{\rm{IR}}}\right]
+\mathcal{O}(\varepsilon_{\rm{IR}})
+\mathrm{finite\ terms},
\eeq
where $\mathcal{M}^{\rm V}_{\rm{block}}$ and $\mathcal{M}^{\rm B}_{\rm{block}}$ 
are arrays of matrix elements corresponding to the
virtual and Born-level QCD building blocks,
\mbox{$\varepsilon_{\rm{IR}}=2/(4-D)$} stands for the IR pole, and
\mbox{$Q^2=-(p_1-p_2)^2=2p_1\cdot p_2$} is two times the scalar
product of the two quark momenta involved in the given building block.
In the example shown in \reffi{polsum2},
$\mathcal{M}^{\rm{B}}_{\rm{block}}$ is built from $A^{\mu\nu}$ and
$D^\mu$, while $\mathcal{M}^{\rm{V}}_{\rm{block}}$ involves
$A^{\mu\nu}_{\rm V}$ or $D^\mu_{\rm V}$ (defined in
\reffi{virtblocks}) instead.  Relation \refeq{virtcheck} has been used
to verify correctness of the IR structure both for the entire virtual
amplitude as well as for each individual building block.

We now provide an overview of the comparisons of the full integrated
cross section for the process
\mbox{$\rp\rp\to \re^+\nu_{\re}\mu^+\nu_{\mu}\rj\rj+X$} with
previously published results.  All results of our calculation have
been produced using Monte Carlo code originally developed for the
calculation of the NLO QCD corrections to
\mbox{$\Pp\Pp\to \PWp\PWm\Pb\bar\Pb+X\to \Pne\Pep\Pmum \bar\Pnmu\Pb\bar\Pb+X$}~\cite{Denner:2012yc}.  
For practical reasons---a built-in generic interface already existed---we
used the tree-level amplitudes generated with \textsc{OpenLoops}
\cite{Cascioli:2011va} after cross-checking them against the ones
obtained in the approach of \refse{method}.  The virtual amplitudes,
on the other hand, are constructed with the methods described in
\refse{method}.

\begin{enumerate}
\item
The first results for NLO QCD corrections to
\mbox{$\rp \rp\to  \re^+\nu_{\re}\mu^+\nu_{\mu}\rj\rj+X$} have been published in
\citere{Jager:2009xx}.  We reproduced the calculation with the same setup
and input parameters. The events have been generated at the
centre-of-mass energy of \mbox{$\sqrt s=14\TeV$}. In the matrix elements
for all partonic processes, we neglected $s$-channel diagrams and
interferences between $t$ and $u$ channels.
The fixed-width scheme has been used to treat the massive
propagators, with the exception of the Higgs couplings where
$M_{\mathrm{Z}}$ and $M_{\mathrm{W}}$ have been kept complex due
to technical reasons, while $\sin\theta_\rw$ and $\cos\theta_\rw$
are real. The factorization and renormalization scales
have been set to \mbox{$\muF=\muR=M_{\mathrm W}$}.  The values of the VBF
cuts are taken from \citere{Jager:2009xx}, however, the
requirement that the charged leptons fall between the tagging jets
in rapidity (\ref{rapjetslep}) has been omitted.%
\footnote{Private communication during comparisons revealed that
in the results presented in \citere{Jager:2009xx} this cut has been
omitted. A corrected version of the article can be found at
\href{http://arxiv.org/abs/0907.0580}{arXiv:hep-ph/0907.0580}.}

\begin{table}
\renewcommand*\arraystretch{1.3}
\centering
\vspace*{3ex}
\begin{tabular}{cccc}
\hline
PDF set & $\sigma[\fb]$ & $\sigma_{\textsc{JOZ}}[\fb]$ & $\delta$[\%]\\
\hline
\mc{4}{l}{Leading order}  \\
\hline
CTEQ6L1 & 1.4746(7) & 1.478 & $-0.23(5)$\\
MSTW08  & 1.4061(7) & 1.409 & $-0.21(5)$\\
\hline
\mc{4}{l}{Next-to-leading order}  \\
\hline
CTEQ6M & 1.405(1) &  1.404 & $+0.10(9)$\\
MSTW08 & 1.372(1) &  1.372 & $-0.00(9)$\\
\hline
\end{tabular}
\vspace*{3ex}
\caption{Comparison of the integrated cross section $\sigma$ with the
cross section $\sigma_{\textsc{JOZ}}$ presented in \citere{Jager:2009xx}
for the $\mathrm{W}^+\mathrm{W}^+$
production processes at NLO. The error
estimates for $\sigma$ are shown in brackets and affect the last digit of
the result. The statistical error of the cross section
$\sigma_{\textsc{JOZ}}$ is stated to be at the sub-per-mille
level and is not taken into account in the last column.}
\label{Zepcomparison}
\end{table}

The LO and NLO results for the two
PDF sets used in \citere{Jager:2009xx} are shown in \refta{Zepcomparison}.
For both LO and NLO cross sections, the relative
deviation between the results of the two calculations is only 
$\sim0.2$\% or even smaller. These small discrepancies could be
attributed to the slight differences in applying the width scheme (see
above).  However, assuming a statistical error of the results of
\citere{Jager:2009xx} of per-mille (it is stated to be at the
sub-per-mille level), the difference amounts to only $2\sigma$ and is
thus acceptable.  The differences between the two PDF sets are at the
level of 5\% at LO and of 2\% at NLO.
\item 
In \citere{Jager:2011ms} the results for
\mbox{$\rp \rp\to  \re^+\nu_{\re}\mu^+\nu_{\mu} \rj\rj+X$}
have been presented at the
centre-of-mass energy of \mbox{$\sqrt s=7\TeV$}.  While the main focus of
\citere{Jager:2011ms} lies on the inclusion of parton-shower effects,
the NLO QCD result for the cross section is also shown.

As in the previous case, we have reproduced the computation of the
cross section with the same setup, parameters, and
kinematic cuts.  The factorization and renormalization scales
have been set to a dynamic value defined as 
\beqa\nn
&\muR=\muF=\displaystyle \f{p_{\rm T,\mathrm{j}_1}+p_{\rm T,\mathrm{j}_2}+E_{\rm T,W_1}+E_{\rm T,W_2}}{2}\quad
\mathrm{with } &E_{\rm T,W_{1/2}}=\sqrt{M_\mathrm{W}^2+p_{\rm T,W_{1/2}}^2}.\\
\eeqa 
Here, $p_{\rm T,W_{1/2}}$ represents the transverse momentum
of the respective same-flavour lepton--neutrino pair, and 
$p_{\rm T,\mathrm{j}_i}$ are the transverse momenta of the two tagging
jets.  This choice of scale is slightly different from the one in
\citere{Jager:2011ms} where, as required by \textsc{ POWHEG}, the
jets of the underlying Born process were used.  Another difference
between the two calculations lies in the width scheme; while our
calculation used the complex-mass scheme, the results in
\citere{Jager:2011ms} have been obtained using the fixed-width scheme.
The impact of the different scheme choices, however, is known
not to exceed the level of a few per-mille here.

The results for the total NLO cross section shown in
\refta{Jagcomparison} for the two calculations differ by
2.3\%. Considering the small differences in the scale choice, and in particular
the statistical error of $\sigma_{\textsc{JZ}}$, the level of agreement is fully acceptable.

\begin{table}[!htb]
\renewcommand*\arraystretch{1.3}
\centering
\vspace*{3ex}
\begin{tabular}{cccc}
\hline
PDF set & $\sigma[\fb]$ & $\sigma_{\textsc{JZ}}[\fb]$ & $\delta$[\%]\\
\hline
\mc{4}{l}{Leading order}  \\
\hline
MSTW08  & 0.16836(8) & / & /\\
\hline
\mc{4}{l}{Next-to-leading order}  \\
\hline
MSTW08  &  0.1961(2) &  0.201(3) & -2.3(1.5)\\
\hline
\end{tabular}
\vspace*{3ex}
\caption{Comparison of the integrated cross section $\sigma$ with
the cross section $\sigma_{\textsc{JZ}}$ presented in \citere{Jager:2011ms}
for the $\mathrm{W}^+\mathrm{W}^+$ production processes NLO. The error
estimates are shown in brackets and affect the last digit(s) of
the result.}
\label{Jagcomparison}
\end{table}

\end{enumerate}

\section{Numerical results}\label{chaptresults}

\subsection{Input parameters and setup}\label{input}
All EW Standard Model parameters used in the calculation are
determined from the values of the Z-boson mass $M_{\rm Z}$, the W-boson mass
$M_{\mathrm{W}}$, the Higgs-boson mass $M_{\rm H}$, and the Fermi coupling
constant $\GF$~\cite{Beringer:1900zz}.
The EW mixing angle $\theta_{\mathrm{w}}$  is defined as
\beq
\cos\theta_{\mathrm{w}}=\f{M_{\mathrm{W}}}{M_{\rm Z}}.
\eeq
The fine-structure constant $\alpha$ is evaluated from $\GF$,
$M_{\mathrm{W}}$ and $M_{\rZ}$ according to
\beq\label{eq:alpha}
\alpha=\f{\sqrt2 M_{\mathrm{W}}^2 \GF}{\pi}\left(1-\f{M_{\rW}^2}{M_{\rZ}^2}\right),
\eeq
which takes into account dominant effects associated with the running
of $\alpha$ from zero to the W-boson mass and absorbs leading
universal corrections $\propto\GF\Mt^2$ associated with the $\rho$
parameter~\cite{Dittmaier:2001ay}.

For all results presented in this section, we make use of the PDF set
MSTW2008~\cite{Martin:2009iq}, i.e.\ MSTW2008LO and MSTW2008NLO for LO and
NLO cross sections, respectively.  Throughout, the NLO value of the
strong coupling constant $\alpha_\rs$ provided by this PDF set is used
(no strong couplings appear at the LO).

The decay widths of the unstable intermediate vector bosons are
calculated at NLO QCD level according to 
\beqa\label{widths}
\Gamma_{\mathrm{W}}&=&\f{\alpha}{6}M_{\mathrm{W}}\left[3\left(\f{1}{\sqrt 2\sin\theta_{\mathrm{w}}}\right)^2
+2\,N_c\left(\f{1}{\sqrt 2\sin\theta_{\mathrm{w}}}\right)^2
\left(1+\f{\alphas(M_{\rm Z})}{\pi}\right)\right],\nn\\
\Gamma_{\mathrm{Z}}&=&\f{\alpha}{6}M_{\mathrm{Z}}\left[\sum_{l}
\left(\left(-Q_l\f{\cos\theta_{\mathrm{w}}}{\sin\theta_{\mathrm{w}}}\right)^2
+\left(\f{I^3_l}{\cos\theta_{\mathrm{w}}\sin\theta_{\mathrm{w}}}
-Q_l\f{\cos\theta_{\mathrm{w}}}{\sin\theta_{\mathrm{w}}}\right)^2\right)\right.\nn\\
&&{}+\left.N_c\sum_q\left(\left(-Q_q\f{\cos\theta_{\mathrm{w}}}{\sin\theta_{\mathrm{w}}}\right)^2
+\left(\f{I^3_q}{\cos\theta_{\mathrm{w}}\sin\theta_{\mathrm{w}}}
-Q_q\f{\cos\theta_{\mathrm{w}}}{\sin\theta_{\mathrm{w}}}
\right)^2 \right) \left(1+\f{\alpha_\rs(M_{\rm Z})}{\pi}\right)\right],\nn\\
\eeqa 
where $l$ runs over all charged leptons and neutrinos, $q$ runs over the
five light quarks, $N_c=3$ is the number of quark colours, and $Q_l$, $Q_q$ and $I^3_l$, $I^3_q$ are the
charges and third isospin components of the respective leptons and
quarks. As the leptonic decays of the EW bosons do not receive QCD
corrections at NLO, we may use the same NLO values of the widths,
provided by \refeq{widths}, both at LO and NLO without introducing
inconsistencies in $\PW$ or $\PZ$ branching ratios.

Throughout the subsequent numerical discussion, we evaluate cross
sections and distributions to
\mbox{$\rp\rp\to \re^+\nu_{\re}\mu^+\nu_{\mu}\rj\rj+X$} at the
centre-of-mass energy \mbox{$\sqrt s=14\TeV$}, using the following Standard
Model parameters,
\begin{alignat}{3}
M_\mathrm{W}&=80.399{\GeV},\qquad&\Gamma_\mathrm W&=2.099736097449861\GeV,\nn\\
M_\mathrm{Z}&=91.1876\GeV,\qquad&\Gamma_\mathrm Z&=2.509659634331562\GeV,\nn\\
M_\mathrm{H}&=125\GeV,\qquad&\Gamma_\mathrm H&=4.07\times10^{-3}\GeV\phantom{0000000}\nn\\
G_{\rm F}   &=1.16637\times10^{-5}\GeV^{-2},\qquad\phantom{00}&\alpha_\rs(M_\mathrm W)&=0.1225519862138941,\label{finalparam}
\end{alignat}
where $\alpha_\rs(M_\mathrm W)$, $\Gamma_\mathrm W$, and $\Gamma_\mathrm Z$ are calculated
values and thus stated at machine precision to facilitate comparisons with our results.
The decay width of the Higgs boson $\Gamma_{\mathrm{H}}$ depends on
the chosen mass, and its value is taken from \citere{Dittmaier:2012vm}.

To treat the propagators of the unstable massive intermediate
particles (W, Z, and Higgs boson), we use the complex-mass scheme
\cite{Denner:1999gp,Denner:2005fg,Denner:2006ic}, in which the masses are globally replaced according to
\beq\label{cms}
M_{V}^{\rm{CMS}}=\sqrt{M_V^2-\ri M_V\Gamma_V}.
\eeq
Complex masses are then introduced everywhere in the Feynman rules,
including the weak mixing angle,
\beq\label{cmstheta}
\cos^2\theta_{\mathrm{w}}\to\f{M_{\mathrm{W}}^2-\ri M_{\mathrm{W}}\Gamma_\mathrm{W}}{M_\mathrm{Z}^2-\ri M_\mathrm{Z}\Gamma_\mathrm{Z}},
\eeq
rendering the couplings complex. Note, that real masses and mixing
angle are used to determine the input values \refeqs{eq:alpha} and
\refeq{widths}.

The LO cross section has been evaluated for three different setups. In
the first, we only take into account the $t$-channel and $u$-channel
diagrams and completely disregard the interferences between them.
This approximation corresponds to the setup in \citere{Jager:2009xx}
and is referred to as VBF approximation.  In the second setup we
include interferences between $t$-channel and $u$-channel diagrams,
and in the third one we calculate the complete cross section including
$t$-, $u$- and $s$-channel diagrams and all interferences. This allows
to assess the size of the $s$-channel and interference contributions.
For the NLO cross section, the $s$-channel diagrams and interferences
are neglected throughout, both in virtual and real corrections.

As in the comparisons of integrated results in the previous section,
all cross sections and distributions have been produced using Monte
Carlo code developed for the calculation of the NLO QCD corrections to
\mbox{$\Pp\Pp\to \PWp\PWm\Pb\bar\Pb+X\to \Pne\Pep\Pmum\bar\nu_\mu\Pb\bar\Pb+X $}~\cite{Denner:2012yc}, 
using adapted tree-level amplitudes generated with \textsc{OpenLoops}
\cite{Cascioli:2011va} while the virtual corrections were calculated
according to the method described in \refse{method}.

\subsection{Jet recombination and phase-space cuts}\label{cuts}

In order to enhance regions of the phase-space where VBF-type
processes can be observed experimentally and QCD background is
reduced, a number of kinematic cuts has been imposed at the Monte
Carlo level.  These cuts are selected to focus on the phase-space
region dominated by VBF processes, which typically contain two hard
jets with large rapidity separation, the so-called tagging jets, and
most of the decay products of the vector bosons in the central detector region.
Further, a set of lepton cuts is applied to ensure that the charged
leptons, which define the respective final state, are well-observable
and separated from the jet activity.  The set of cuts to be precisely
defined in this section  follows the proposal of
\citere{Jager:2009xx}.

To be considered protojets which eventually give rise to hadronic jets in 
the final state, outgoing QCD partons have to fulfil the requirement
\beq
|\eta|=\left|\f{1}{2}\ln\f{p_0+p_z}{p_0-p_z}\right|\leq5.
\eeq
Starting with these protojets, the jet reconstruction is performed
using the $k_{\rT}$ algorithm~\cite{Catani:1992zp,Blazey:2000qt} with
the resolution parameter \mbox{$D=0.7$}.  In order to be clearly
distinguished from QCD background, the resulting jets must satisfy the
transverse momentum and rapidity cuts
\beq
p_{\mathrm{T,j}}=\sqrt{p_{x,\rj}^2+p_{y,\rj}^2}\geq20{\GeV,}\qquad
|y_\rj|=\left|\f{1}{2}\ln\f{p_{0,\rj}+p_{z,\rj}}{p_{0,\rj}-p_{z,\rj}}\right|\leq4.5.
\eeq
At least two jets have to pass this criterion, and the two jets with
the highest transverse momenta are denoted as tagging jets, on which
the following additional restrictions are imposed.
The two tagging jets must have a minimum invariant mass,
\beq\label{mjj}
M_{\mathrm{\rj\rj}}=\sqrt{(p_{0,\rj_1}+p_{0,\rj_2})^2-(\vec{\mathbf{p}}_{\rj_1}+\vec{\mathbf {p}}_{\rj_2})^2}>600{\GeV},
\eeq
they must be located in the opposite hemispheres of the detector,
\beq
y_{\rj_1}\times y_{\rj_2}<0,
\eeq
and show a large rapidity separation,
\beq\label{rapsep}
\Delta y_{\mathrm{\rj\rj}}=|y_{\rj_1}-y_{\rj_2}|>4,
\eeq
in order to further suppress the gluon-induced production mode and
background processes.

The charged leptons are required to pass transverse-momentum and
rapidity cuts,
\beq\label{ptlep}
p_{\mathrm{T,l}}\geq20{\GeV,}\qquad
|y_{\mathrm{l}}|\leq 2.5.
\eeq
To ensure that they are well separated from one another and from the
two tagging jets, we impose the additional cuts
\beq\label{deltaRjl}
\Delta R_{\mathrm{jl}}\geq0.4,\qquad
\Delta R_{\mathrm{ll}}\geq0.1,
\eeq
where the quantity $\Delta R_{ij}$ is a measure of distance in rapidity and azimuthal angle, defined as
 \beq\label{separation}
\Delta R_{ij}=\sqrt{(y_i-y_j)^2+(\phi_i-\phi_j)^2},
\eeq
where $y_i$, $y_j$ and $\phi_i$, $\phi_j$ are the rapidities and
azimuthal angles of the respective particles.  Finally, the rapidities
of the charged leptons are required to fall between the tagging-jet
rapidities,
\beq\label{rapjetslep}
y_{\mathrm{j_{min}}}<y_\mathrm{l}<y_{\mathrm{j_{max}}},
\eeq
which again points out a typical feature of the VBF-production mode.

\subsection{Integrated cross sections}

We have chosen two types of scales to demonstrate their effects on
the behaviour of the NLO distributions for selected observable
quantities. In the fixed-scale (FS) choice, both factorization and
renormalization scales have been set to the mass of the W boson, which
sets a natural scale for the total cross section of the process
\mbox{$\Pp\Pp\to \PWp\PWp \rj\rj+X\to \Pep\Pne\mu^+\nu_\mu \rj\rj+X$}, 
and varied by a factor $\xi$ around this central value,
\beq\label{FS1}
\muF=\muR=\xi M_{\mathrm W}.
\eeq
Since this FS choice turns out to result in strongly phase-space dependent $K$ factors---%
in particular in the high-energy tails of distributions---a dynamical
scale (DS),
\beq\label{DS}
\muF=\muR=\xi \sqrt{p_{\mathrm{T,j_1}}\cdot p_{\mathrm{T,j_2}}}.
\eeq
has been considered as well. Unlike the DS chosen in \citere{Jager:2009xx}, \refeq{DS}
only depends on final-state momenta and can thus be easily defined in an IR-safe way
also at NLO.
The scale in \refeq{DS} has been chosen to flatten the variation of the $K$ factor
in the high-energy tails of $p_{\mathrm{T,j}}$ as well as of other energy-dependent
distributions, which is demonstrated in the next section.

\begin{figure}
\centering
\subfloat[$\muF=\muR=\xi M_{\mathrm W}$]
{\label{FS}\epsfig{figure=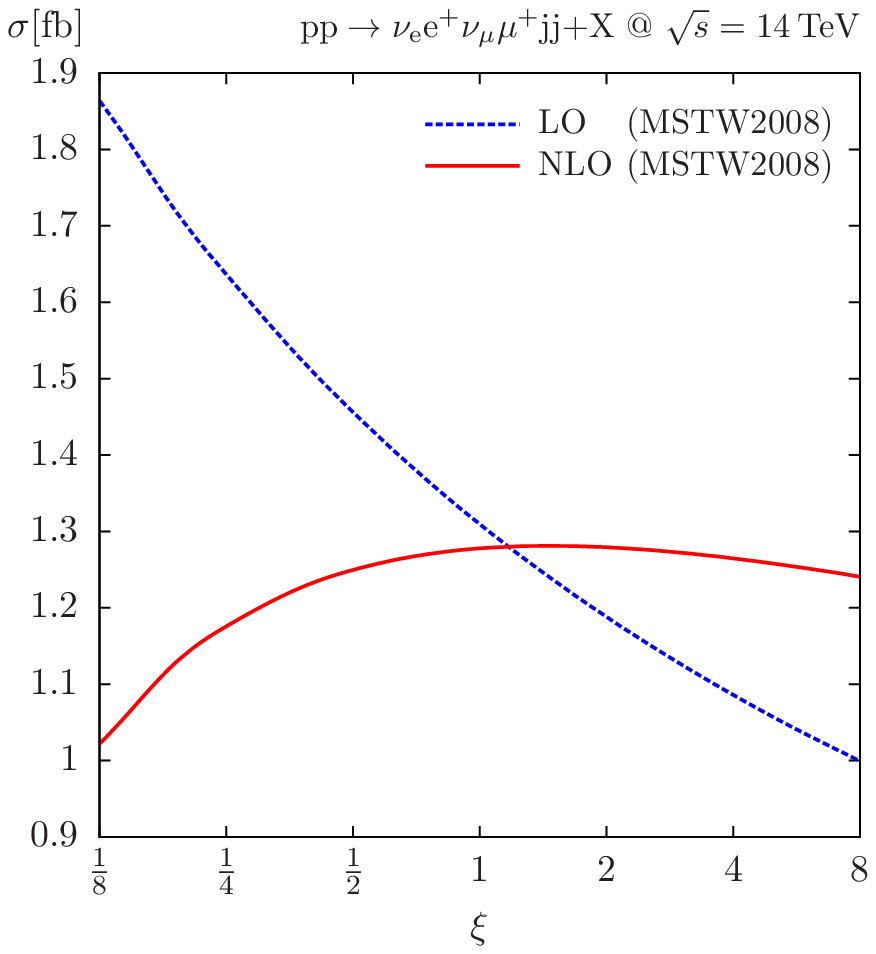,width=7cm}}\qquad
\subfloat[$\muF=\muR=\xi \sqrt{p_{\mathrm{T,j_1}}\cdot p_{\mathrm{t,j_2}}}$]
{\label{DS2}\epsfig{figure=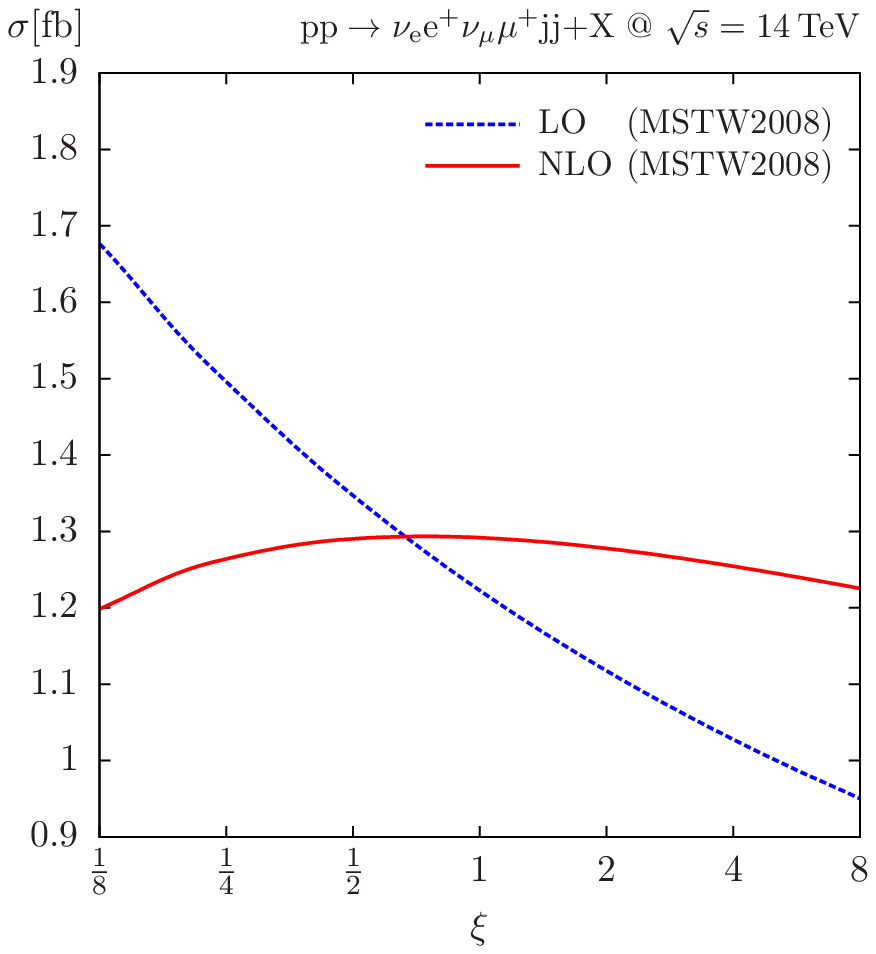,width=7cm}}
\caption{Scale dependence of the LO (dotted blue line) and NLO (solid
red line) cross section for the fixed (\reffi{FS}) and 
dynamic scale (\reffi{DS2}) as a function of the scale parameter $\xi$.}
\label{scales}
\end{figure}

The dependence of the total cross section on the parameter $\xi$ for
both scale choices is depicted in \reffi{scales} for a variation
of $\xi$ in the range \mbox{${1}/{8}<\xi<8$}. In the conventionally chosen
range \mbox{${1}/{2}<\xi<2$}, the scale variation of the LO cross section
which only depends on $\xi$ via $\muF$ (as $\muR$ only enters in
$\alpha_\rs(\muR)$) is about $\pm10\%$,
while at NLO it is reduced to about $\pm2\%$ of the total cross
section for the FS choice and $\pm1\%$ for the DS choice.  For scales
down to $\xi=1/8$, the scale dependence for the DS is less pronounced
($\sim-7\%$ of the total cross section at $\xi=1$) than in the case of
the FS ($\sim-20\%$ of the total cross section at $\xi=1$).  For both
scale choices the maximum of the NLO curve is located in the vicinity
of \mbox{$\xi=1$} (reflecting a small residual scale dependence in
this region), so this value is chosen for the subsequent numerical
discussions.  For \mbox{$\xi=1$}, the overall $K$ factor, defined as
\mbox{$\sigma^\mathrm{NLO}/\sigma^\mathrm{LO}$}, is
\beq\label{Kfactorres}
K_{\rm{FS}}=0.976,\qquad K_{\rm{DS}}=1.056,
\eeq
respectively, for the two scale choices under consideration.\\

\begin{table}
\renewcommand*\arraystretch{1.3}
\centering
\begin{tabular}{c|ccc|c}
\hline
$\xi$ & $\sigma^{\mathrm{LO}}_{\mathrm{full}}[\fb]$ & $\sigma^{\mathrm{LO}}_{\mathrm{VBF+int}}[\fb]$ & $\sigma^{\mathrm{LO}}_{\mathrm{VBF}}[\fb]$ & $\sigma^{\mathrm{NLO}}_{\mathrm{VBF}}[\fb]$ \\
\hline
1/8      &   1.6763(2) &       1.6755(2) &       1.6771(2) &     1.198(2) \\
1/4      &   1.4956(2) &       1.4949(2) &       1.4964(2) &     1.264(1) \\
1/2      &   1.3467(2) &       1.3461(2) &       1.3474(2) &    1.2903(9) \\
1        &   1.2224(2) &       1.2218(2) &       1.2230(2) &    1.2917(8) \\
2        &   1.1173(2) &       1.1168(2) &       1.1179(2) &    1.2778(7) \\
4        &   1.0275(2) &       1.0270(2) &       1.0280(2) &    1.2544(6) \\
8        &   0.9499(2) &       0.9494(2) &       0.9504(2) &    1.2253(6) \\
\hline
\end{tabular}
\vspace*{3ex}
\caption{Integrated cross sections for LO including all channels and interferences ($\sigma^{\mathrm{LO}}_{\mathrm{full}}$),
for LO including $t$--$u$ interferences but neglecting $s$-channel diagrams ($\sigma^{\mathrm{LO}}_{\mathrm{VBF+int}}$),
for LO in the VBF approximation, i.e.\ neglecting all $s$-channel
diagrams and interferences
($\sigma^{\mathrm{LO}}_{\mathrm{VBF}}$), and 
for NLO in the VBF approximation
($\sigma^{\mathrm{NLO}}_{\mathrm{VBF}}$).
The integration-error estimates are shown in brackets and affect
the last digit of the respective result.} 
\label{finalcs}
\end{table}

The dedicated VBF cuts listed in \refse{cuts} prefer
$t$- and $u$-channel kinematics, whereas $s$-channel configurations
are strongly suppressed by the requirement of final-state jets
with large rapidity separation and invariant mass. Moreover,
interferences between $t$ and $u$ channels, showing up in partonic processes
with identical final-state \mbox{(anti-)quarks}, are suppressed by the
condition that the tagging jets have to be located in opposite---forward
and backward---regions of the detector. It can therefore be
argued~\cite{Jager:2009xx} that the $s$-channel and interference contributions
can safely be neglected if the VBF cuts are applied. In order to
verify this claim, the LO cross section has been evaluated
for three different sets of matrix elements; the results
(obtained using the DS with selected values of parameter $\xi$) can be found in
\refta{finalcs}.  Here, $\sigma^{\mathrm{LO}}_{\mathrm{full}}$
stands for the cross section that includes all channels and interferences,
while $\sigma^{\mathrm{LO}}_{\mathrm{VBF+int}}$ contains the complete
$t$- and $u$-channel contributions with interferences but no
$s$-channel contributions, and $\sigma^{\mathrm{LO}}_{\mathrm{VBF}}$
contains only squares of $t$- and $u$-channel contributions but no
interferences. One can see that for the cross section within our set of
VBF cuts the effect of the $t$--$u$ interferences is at the level of
$-0.05\%$, and the contribution of $s$-channel diagrams at the level
of $+0.1\%$. This confirms that $\sigma^{\mathrm{LO}}_{\mathrm{VBF}}$ can
be considered a very good approximation of the full LO cross section.
For this reason, the NLO cross section has been evaluated using only
$t$- and $u$-channel contributions without interferences between them
in order to improve the speed of the calculation.  The values of
$\sigma^{\mathrm{NLO}}_{\mathrm{VBF}}$ for different values of $\xi$
can be found in the fifth column in \refta{finalcs}.

\subsection{Jet distributions}

All distributions shown in this and the following section are
evaluated in the numerical setup of \refse{input}, using the kinematic
cuts introduced in \refse{cuts}; the scale choice applied is stated in
each case.

Two plots are presented for each observable: the one on the left
depicts the LO and NLO predictions, the uncertainty of which is
indicated by error bands resulting from variation of the given scale
within \mbox{$1/2<\xi<2$}, while the plot on the right shows the LO
and NLO predictions normalized to the LO result at the central scale,
i.e.\ \mbox{$K_{\mathrm{LO}}(\xi)=\rd \sigma_{\mathrm{LO}}(\xi)/\rd\sigma_{\mathrm{LO}}(\xi=1)$} 
(dotted blue line), and 
\mbox{$K_{\mathrm{NLO}}(\xi)=\rd \sigma_{\mathrm{NLO}}(\xi)/\rd\sigma_{\mathrm{LO}}(\xi=1)$} 
(solid red line). The blue band in this case corresponds to the relative
scale uncertainty of the cross section at LO, and the central curve of
the red band represents the conventional $K$ factor
\mbox{$K_\mathrm{NLO}(\xi=1)$}.

\begin{figure}
\centering 
\subfloat[$\muF=\muR=M_{\mathrm W}$]{\label{pTmaxFS}\epsfig{figure=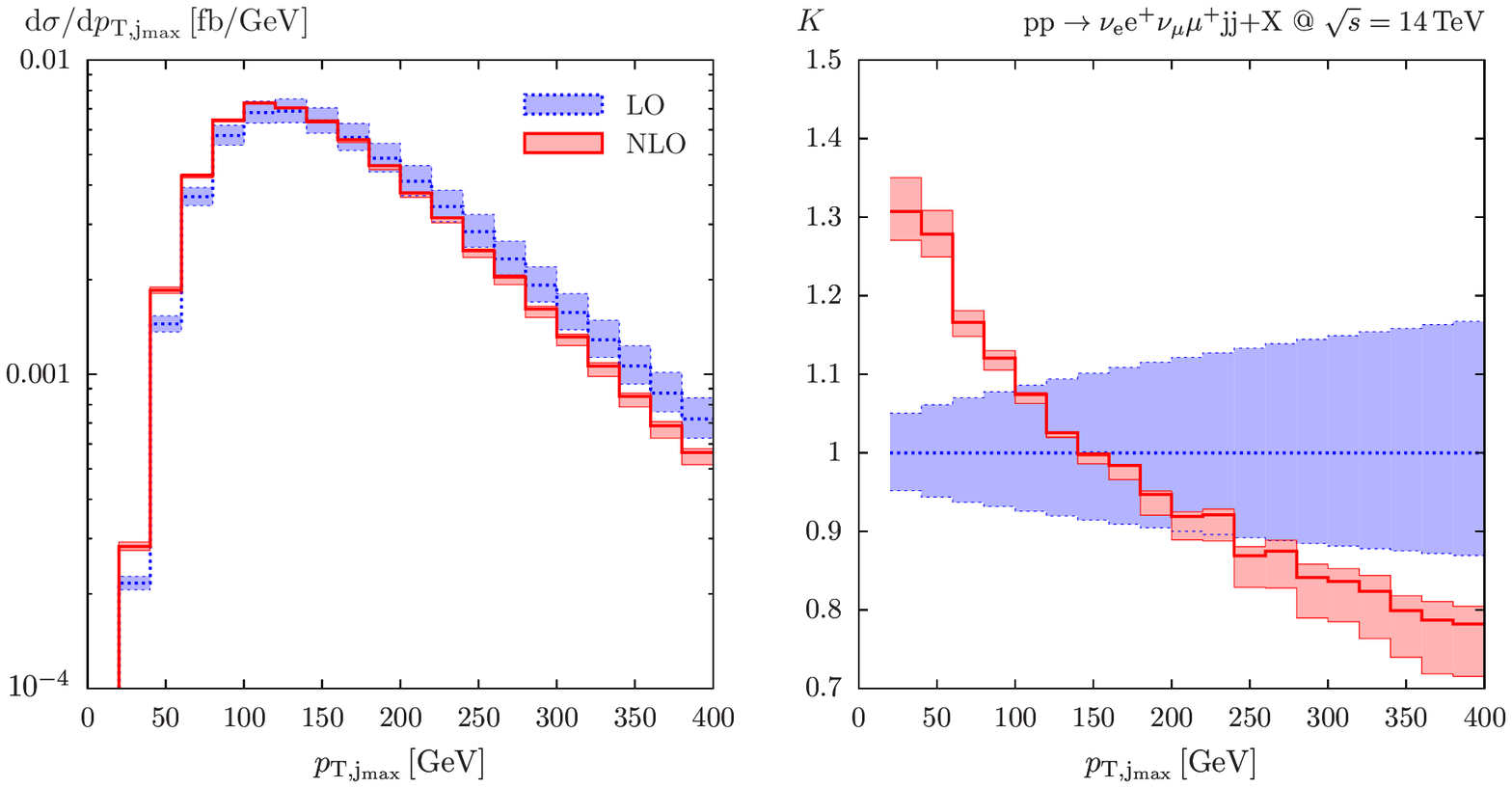,width=14cm}}\\
\subfloat[$\muF=\muR=\sqrt{p_{\mathrm{T,j_1}}\cdot p_{\mathrm{T,j_2}}}$]{\label{pTmaxDS2}\epsfig{figure=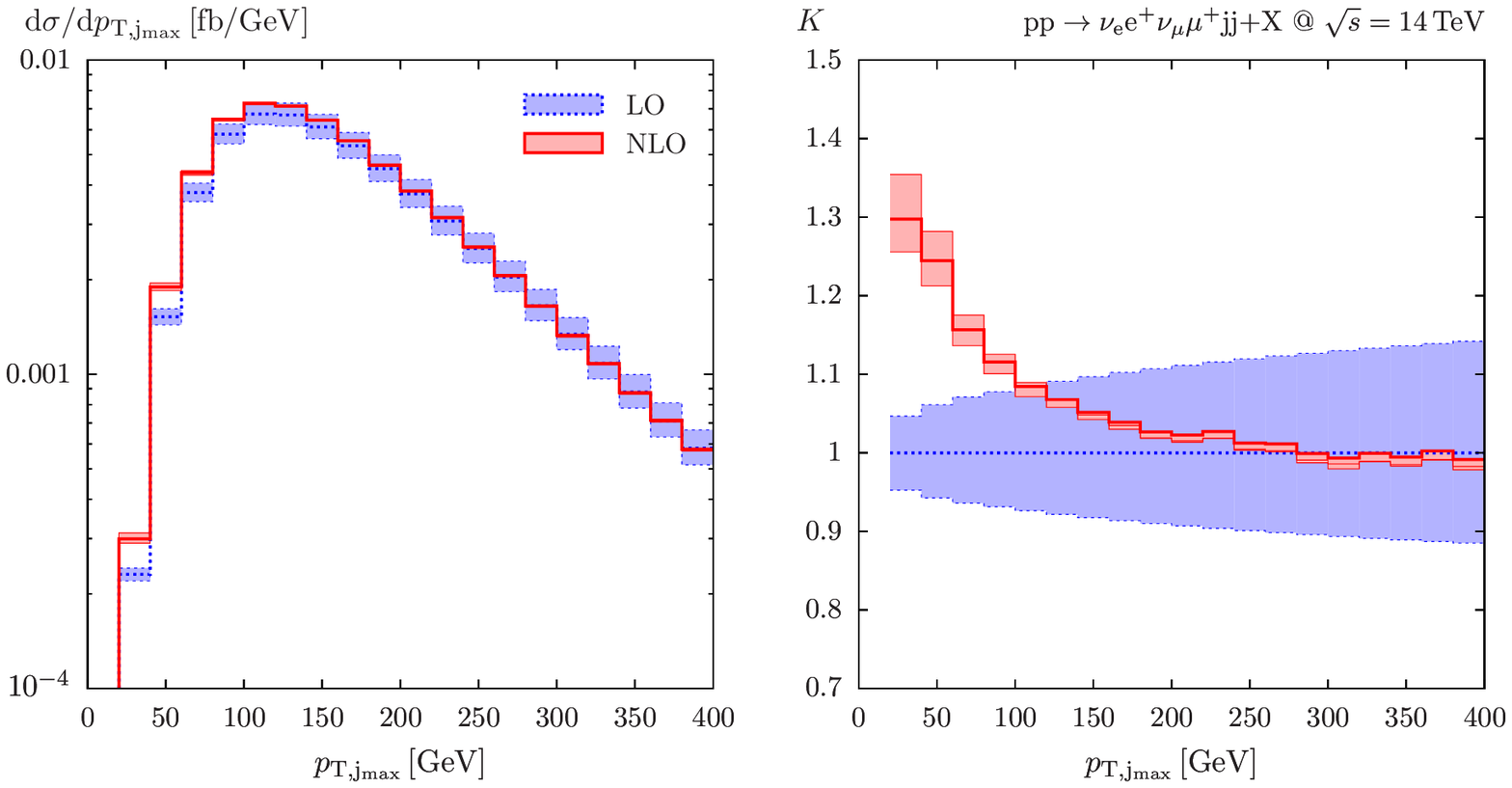,width=14cm}}
\caption{Transverse momentum distribution of the tagging jet with 
the higher $\pT$ for the fixed (\reffi{pTmaxFS}) and
dynamic scale (\reffi{pTmaxDS2}) on the left and the
corresponding $K$ factor represented by the solid (red) line on the
right.}
\label{pTmax plot}
\end{figure}

\reffi{pTmax plot} shows the LO and NLO cross sections as functions of
the transverse momenta of the harder (in terms of $\pT$) of the two
tagging jets in the range \mbox{$p_{\mathrm{T,j_{max}}}\leq 400\GeV$}.
\reffi{pTmaxFS} displays the
dependence for the fixed scale (\ref{FS1}) and \reffi{pTmaxDS2}
for the dynamic scale (\ref{DS}). In both cases, the distribution
peaks at \mbox{$p_{\rT} \sim110\GeV$}, confirming the preference of the high-$\pT$
regions by the VBF tagging jets, while the probability to find a jet
at lower values of $\pT$ is slightly larger at NLO than at LO.
One can observe that $K(p_{\mathrm{T,j_{max}}})$ grows noticeably in
low-$\pT$ regions for both FS and DS towards the value 1.3, while in
the larger-$\pT$ regions it drops to 0.8 in case of the FS
(\reffi{pTmaxFS}) and remains very close to 1 for the DS
(\reffi{pTmaxDS2}), which is a behaviour that motivated the choice of the
DS in the first place.

\begin{figure}
\centering
\subfloat[$\muF=\muR=M_{\mathrm W}$]{\label{pTminFS}\epsfig{figure=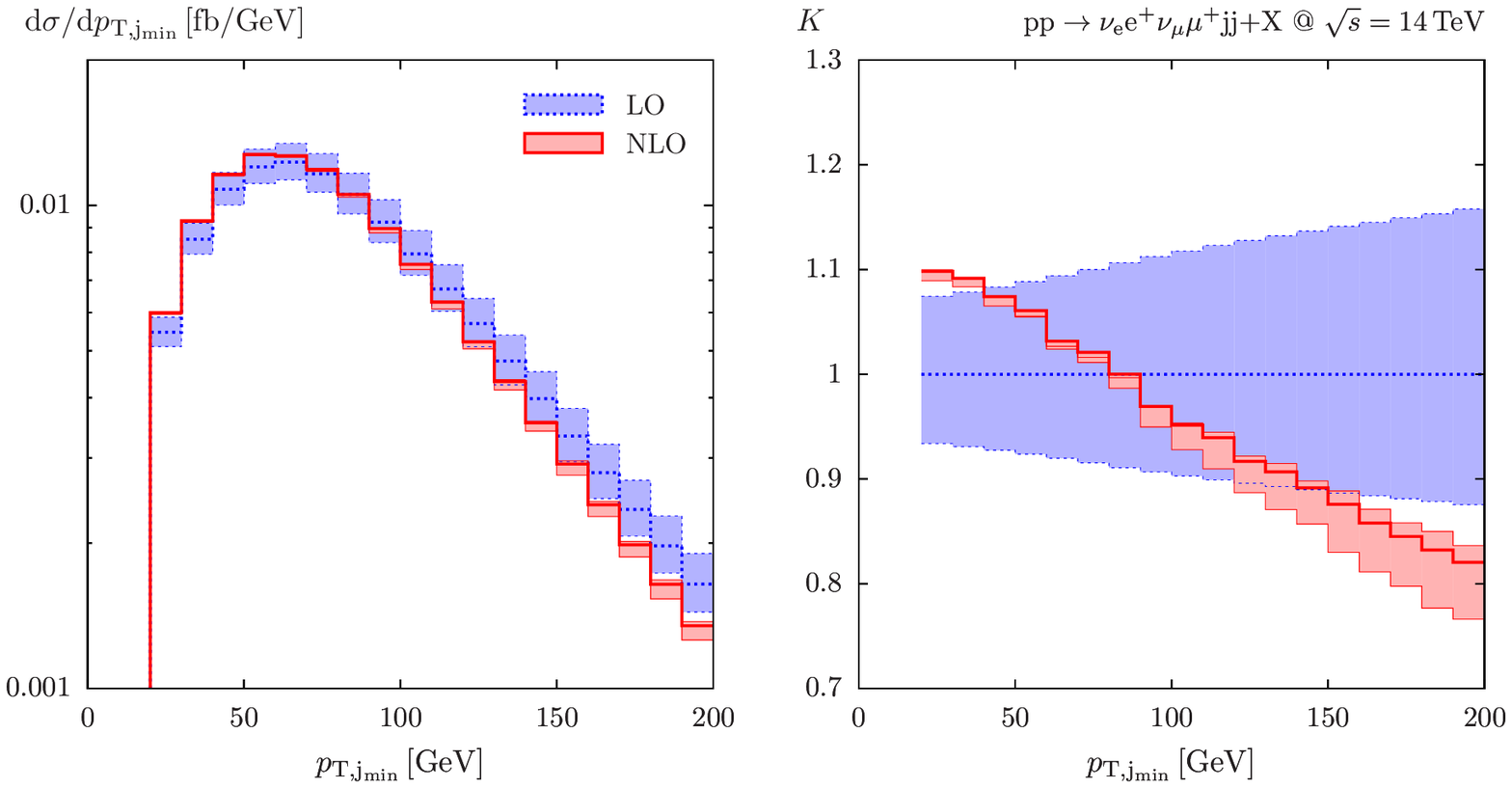,width=14cm}}\\
\subfloat[$\muF=\muR=\sqrt{p_{\mathrm{T,j_1}}\cdot p_{\mathrm{T,j_2}}}$]{\label{pTminDS2}\epsfig{figure=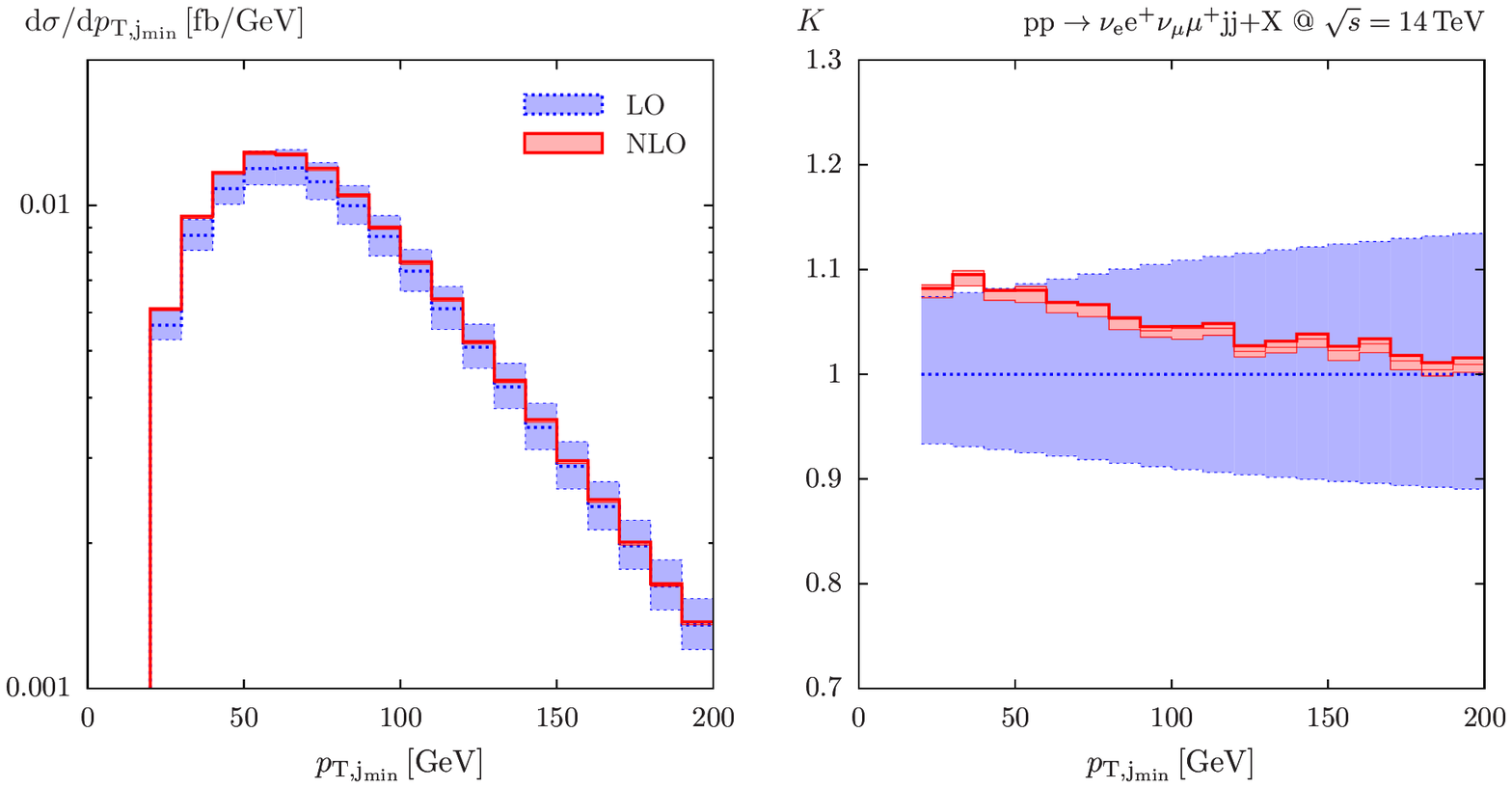,width=14cm}}
\caption{Transverse momentum distribution of the tagging jet with the 
lower $\pT$ for the fixed (\reffi{pTminFS}) and dynamic scale (\reffi{pTminDS2}) 
on the left and the corresponding $K$ factor represented by the solid (red) line 
on the right.}
\label{pTmin plot}
\end{figure}

A similar behaviour can be observed with the transverse momentum of
the softer tagging jet $p_{\mathrm{T,j_{min}}}$, as shown in
\reffi{pTmin plot}. Here, the peak of the distribution is at
\mbox{$\pT\sim 60\GeV$}, indicating that the cut of $20\GeV$ on the
transverse momenta of the tagging
jets does not impose any significant reduction to the overall cross
section. The variation of $K(p_{\mathrm{T,j_{min}}})$ is less
pronounced than in the case of $p_{\mathrm{T,j_{max}}}$, while the
choice of DS again shows an improvement at reducing contributions from
higher-order corrections.

\begin{figure}
\centering
\subfloat[$\muF=\muR=M_{\mathrm W}$]{\label{absyhardFS}\epsfig{figure=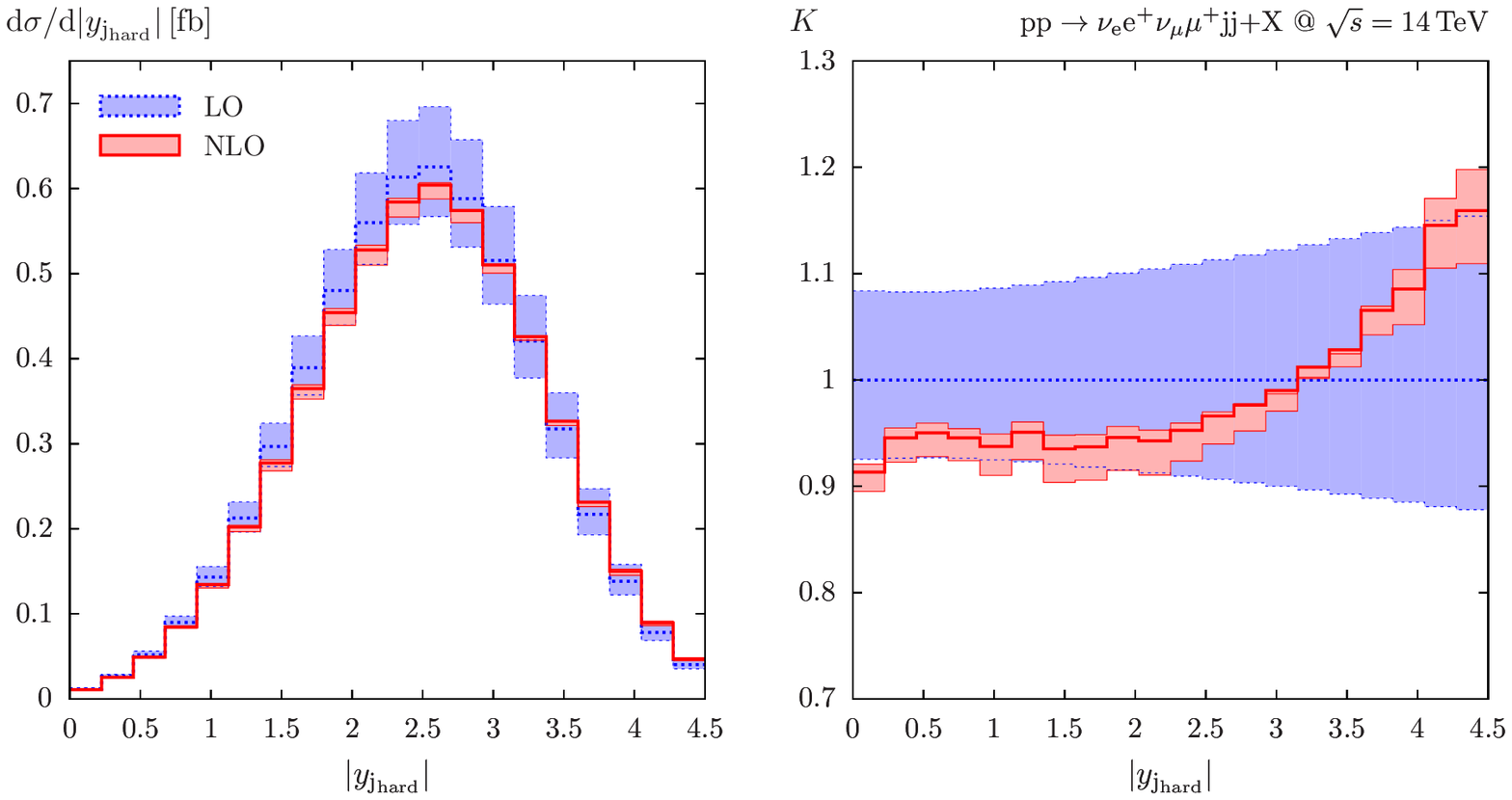,width=14cm}}\\
\subfloat[$\muF=\muR=\sqrt{p_{\mathrm{T,j_1}}\cdot p_{\mathrm{T,j_2}}}$]{\label{absyhardDS2}\epsfig{figure=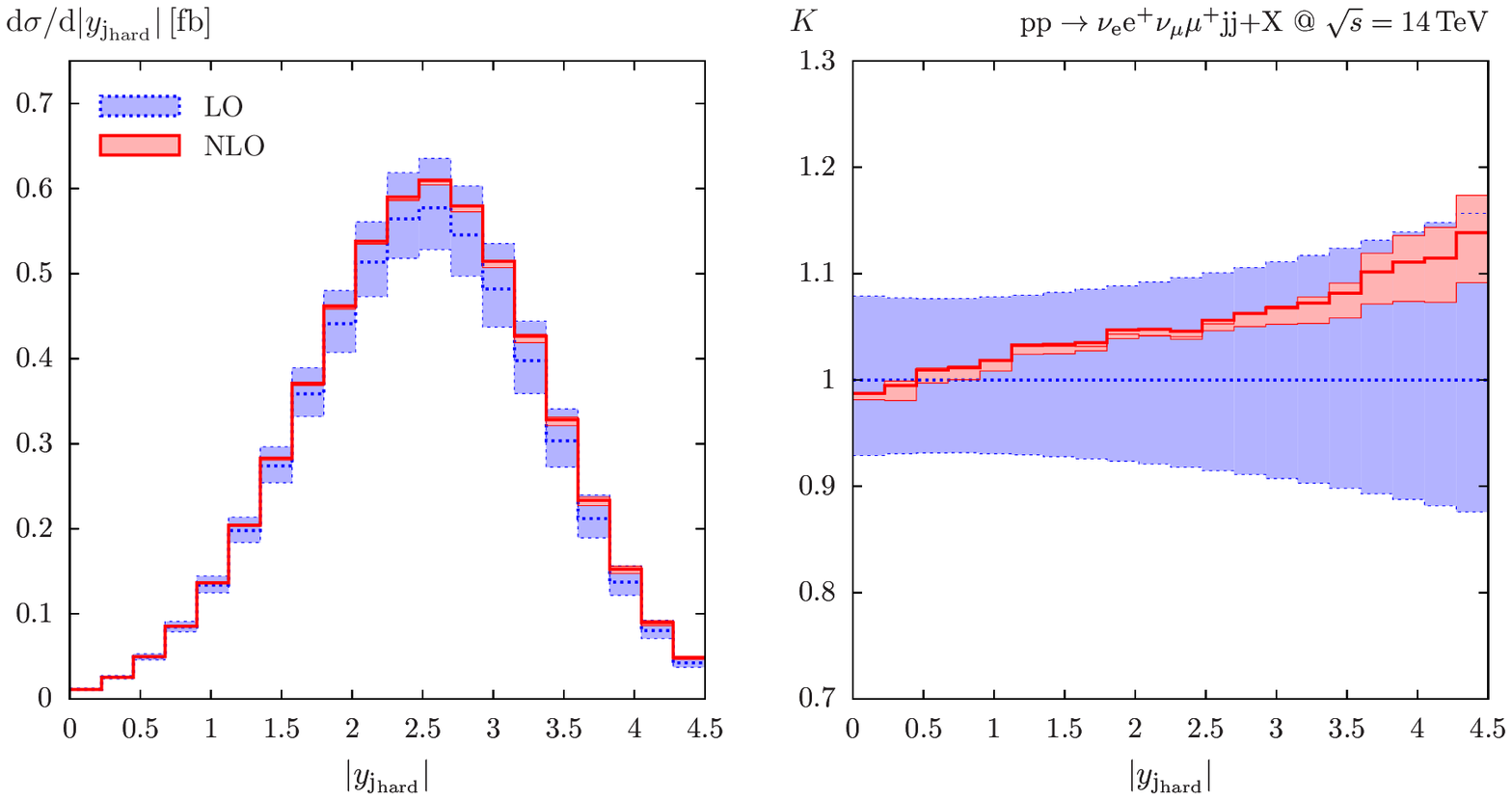,width=14cm}}
\caption{Absolute rapidity distribution for the harder tagging jet for the 
fixed (\reffi{absyhardFS}) and dynamic scale (\reffi{absyhardDS2}) on the left 
and the corresponding $K$ factor represented by the solid (red) line on the right.}
\label{absyhard plot}
\end{figure}

\begin{figure}
\centering
\subfloat[$\muF=\muR=M_{\mathrm W}$]{\label{absysoftFS}\epsfig{figure=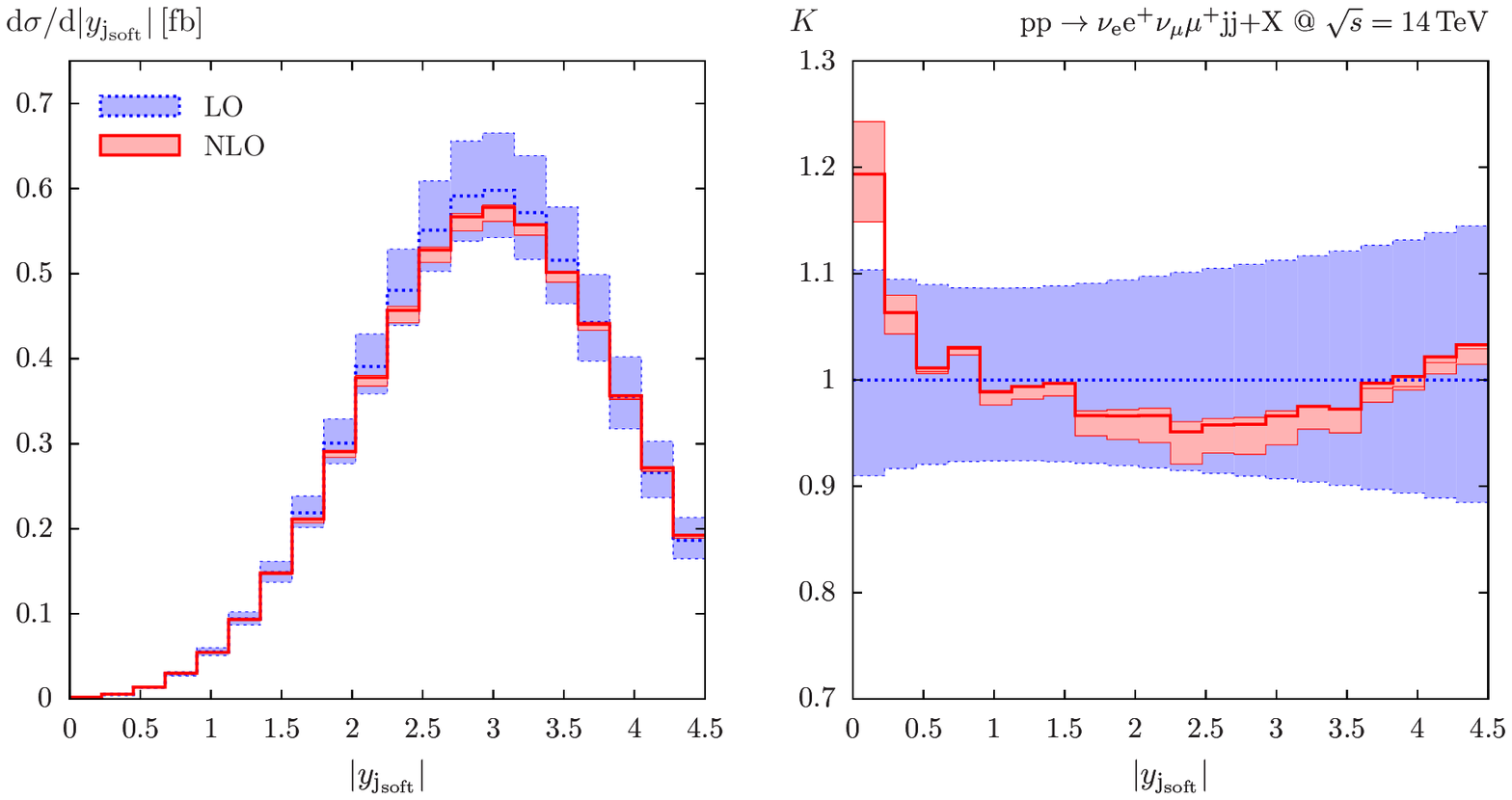,width=14cm}}\\
\subfloat[$\muF=\muR=\sqrt{p_{\mathrm{T,j_1}}\cdot p_{\mathrm{T,j_2}}}$]{\label{absysoftDS2}\epsfig{figure=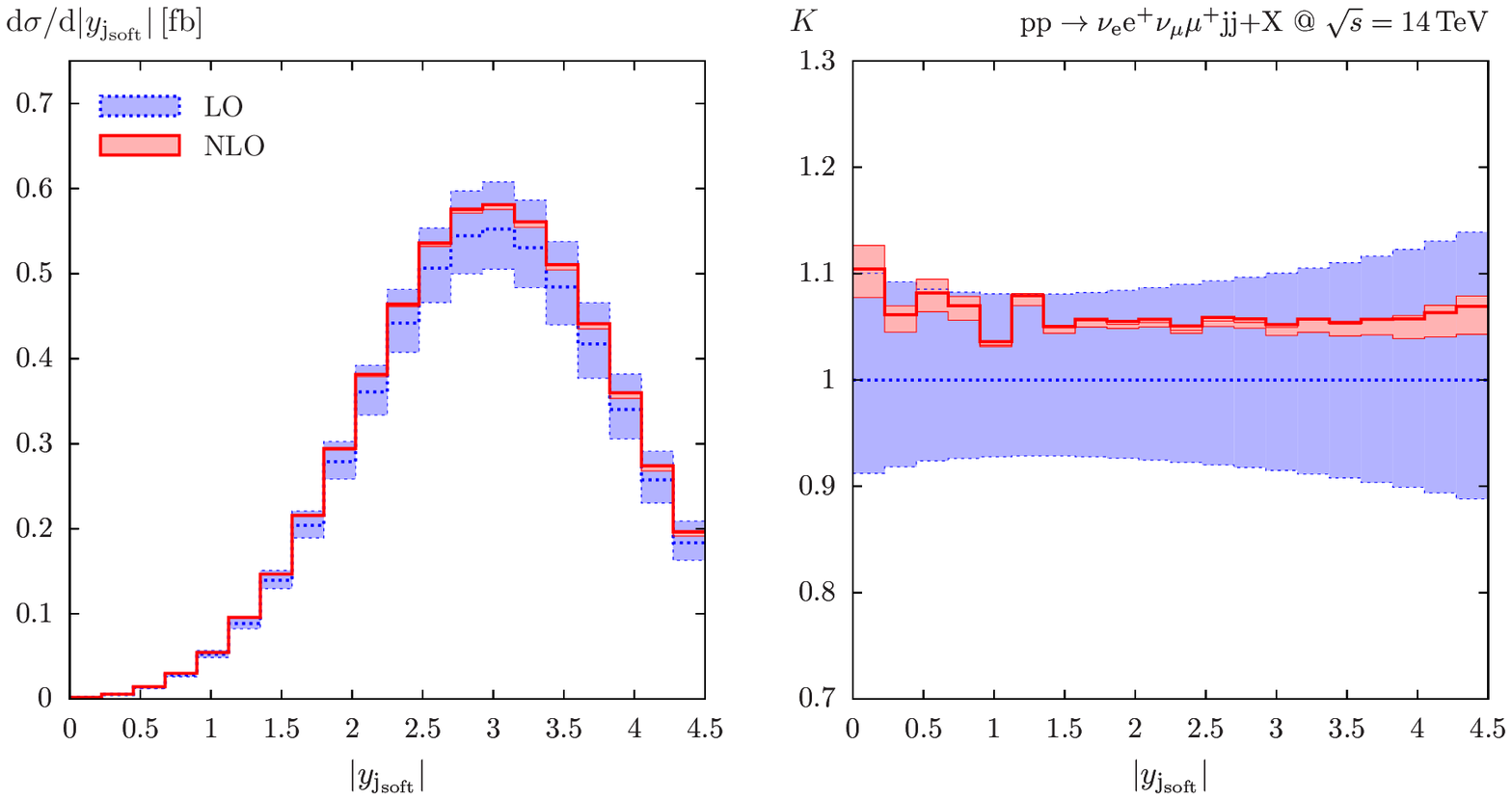,width=14cm}}
\caption{Absolute rapidity distribution for the softer tagging jet $|y_{\mathrm{j_{soft}}}|$ 
for the fixed (\reffi{absysoftFS}) and dynamic scale (\reffi{absysoftDS2}) on the left 
and the corresponding $K$ factor represented by the solid (red) line on the right.}
\label{absysoft plot}
\end{figure}

The rapidity of the tagging jets is another distinguishing feature of
the VBF processes, as they exhibit very little jet activity in the
central region. Absolute rapidity distributions for the harder and
softer (in terms of $\pT$) tagging jets are shown in 
\reffis{absyhard plot}{absysoft plot}, respectively. One can see that the 
probability to find the harder jet peaks at absolute rapidity of \mbox{$y\sim2.6$}
while the softer jet is most likely to be found with absolute rapidity
of \mbox{$y\sim 3.1$}. This is in sharp contrast to the behaviour for
the QCD production mode for $\PWp\PWp$, where the jet rapidity peaks
at 0, dominating the central rapidity region~\cite{Jager:2011ms}.
This production mode thus can be suppressed dramatically by imposing a
cut on the separation of individual jet rapidities $\Delta
y_{\mathrm{\rj\rj}}$ (\ref{rapsep}).  One can see from
\reffis{absyhard plot}{absysoft plot} that for both scale choices the
rapidity-dependent $K$ factor for the hard jet
$K(|y_\mathrm{j_{hard}}|)$ has a tendency to grow for large values of
rapidity. This might be attributed to the fact that while only two
final-state partons are present at LO, in NLO the tagging jets are
selected from up to three partons which might lead to greater
dispersion in the rapidity distribution (see discussion in
\citere{Jager:2009xx}).  As in the case of the transverse-momentum
distributions, the DS shows an improvement over the FS in the
variation of the $K$ factor (\reffis{absyhardDS2}{absysoftDS2}).

\begin{figure}
\centering
\subfloat[$\muF=\muR=M_{\mathrm W}$]{\label{MjjFS}\epsfig{figure=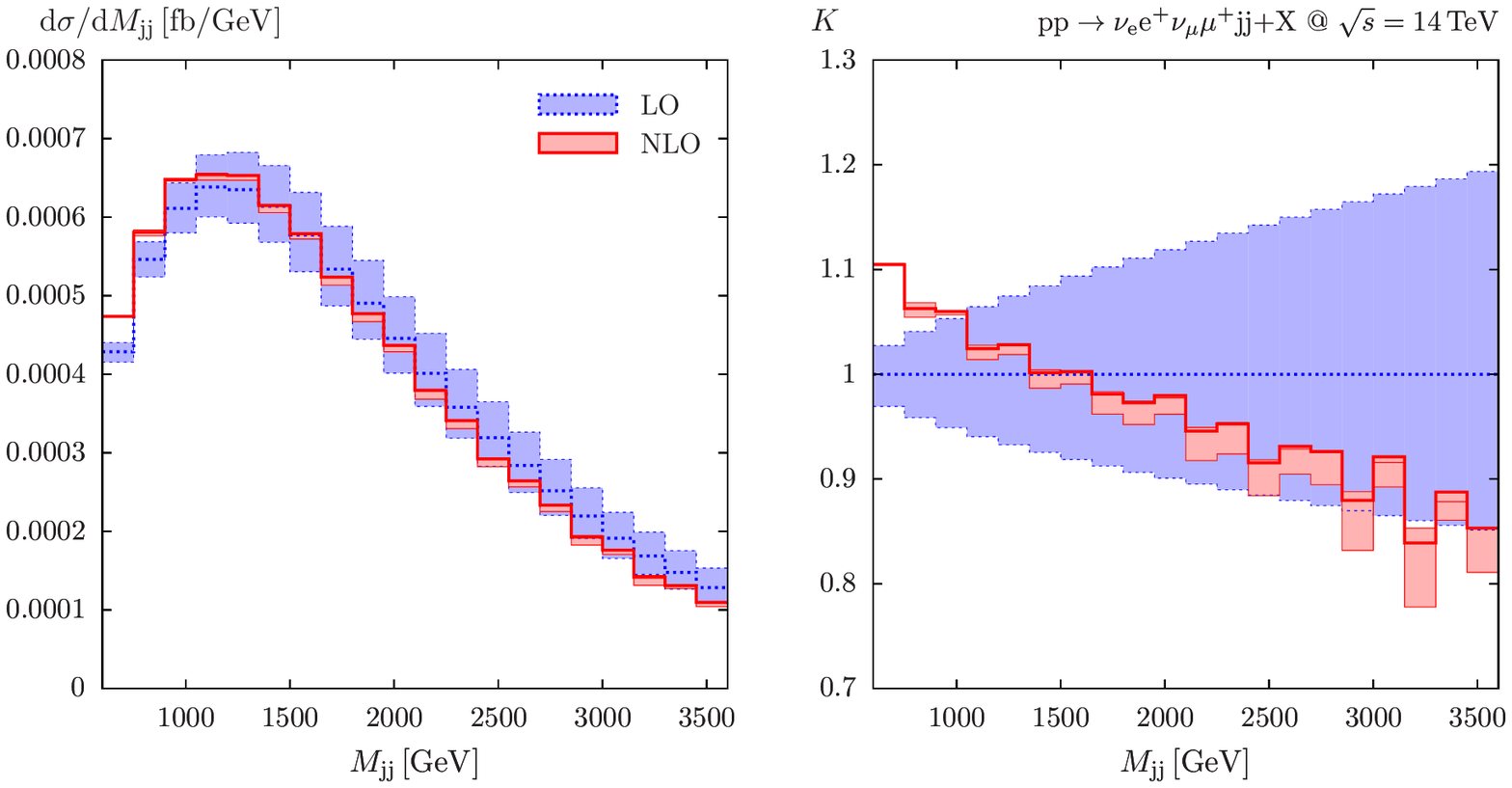,width=14cm}}\\
\subfloat[$\muF=\muR=\sqrt{p_{\mathrm{T,j_1}}\cdot p_{\mathrm{T,j_2}}}$]{\label{MjjDS2}\epsfig{figure=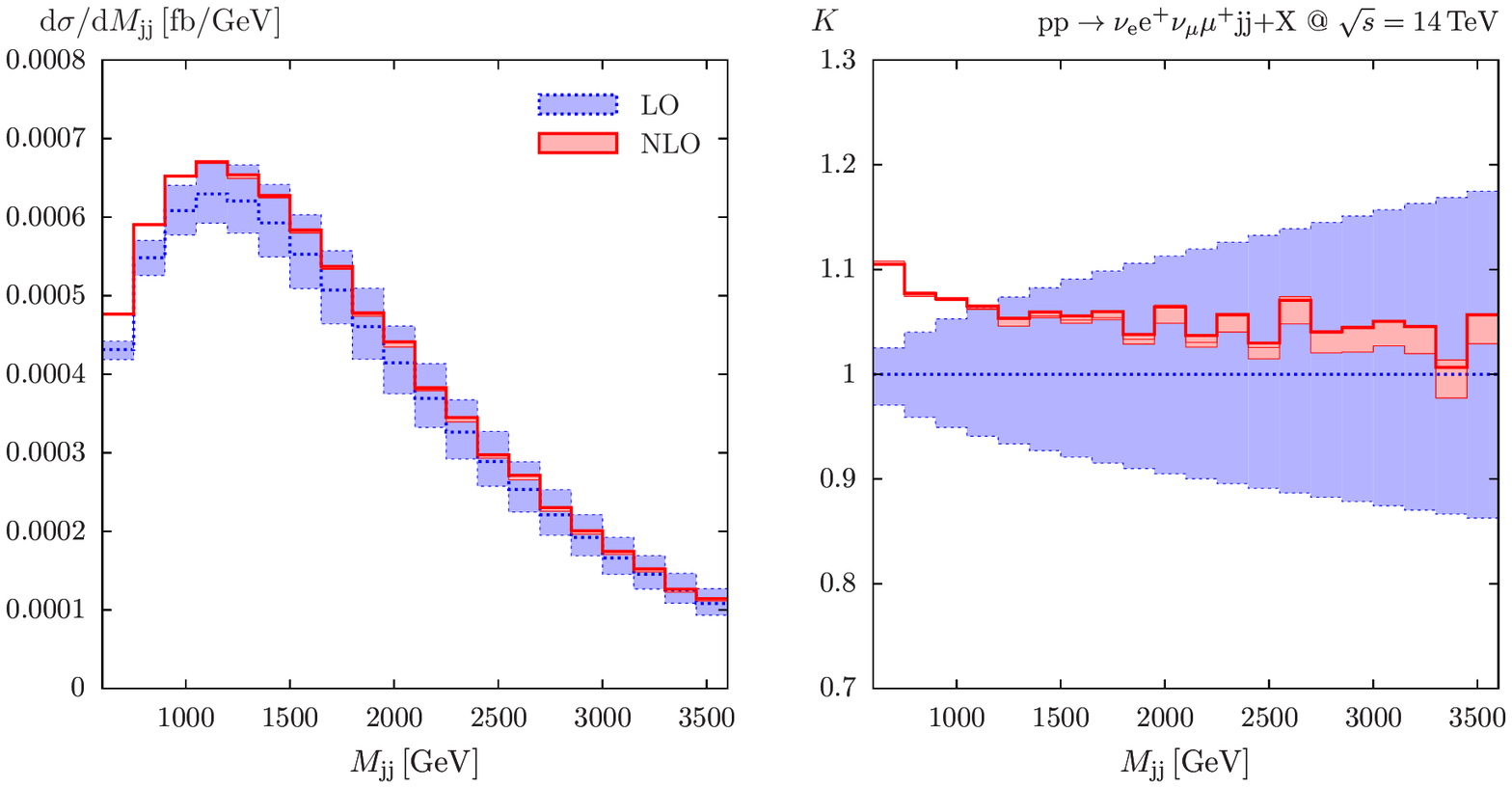,width=14cm}}
\caption{Distribution in the invariant mass of the two tagging jets $M_{\rj\rj}$ 
for the fixed (\reffi{MjjFS}) and dynamic scale (\reffi{MjjDS2}) on the left 
and the corresponding $K$ factor represented by the solid (red) line on the right.}
\label{Mjj plot}
\end{figure}

At hadron colliders, QCD processes typically occur at smaller
energy scales than EW processes. Due to the back-to-back geometry and
large momenta of the tagging jets in VBF, the invariant mass
$M_{\mathrm{\rj\rj}}$ defined in (\ref{mjj}) can easily exceed $1\TeV$,
which is typically not the case for any QCD background process, particularly if
incoming gluons, which prefer smaller momentum fractions than
valence quarks, are involved. For this reason, invariant-mass cuts are applied to
distinguish these types of processes.  \reffi{Mjj plot} provides
the distribution for the invariant mass of the tagging jets for both FS
and DS. The peak of the distribution is located at approximately
$1100\GeV$, both at LO and NLO. On a qualitative level, the behaviour of
the distributions as well as the $K$ factor $K(M_{\rj\rj})$ is in good
correspondence to those shown in \citere{Jager:2009xx}, in particular for
the FS which is set to $M_{\mathrm W}$ in both cases.
While the DS in \citere{Jager:2009xx} is set to the momentum transfer
between the incoming and outgoing partons rather than to
\mbox{$\sqrt{p_{\mathrm{T,j_1}}\cdot p_{\mathrm{T,j_2}}}$}, it has
a similar effect on the behaviour of the NLO distribution, reducing the
scale variation of the $K$ factor.

\subsection{Leptonic distributions}

The decay products of the intermediate gauge bosons in VBF processes
can be found almost exclusively in between the tagging jets, in the
central region of the detector. This kinematic feature is used to
further suppress in particular irreducible background from
gluon-mediated contributions to the process
\mbox{$\Pp\Pp\to \PWp\PWp\rj\rj+X\to \Pep\Pne\mu^+\nu_\mu \rj\rj+X$},
which do not show these characteristics, e.g.\ by imposing
lepton--rapidity cuts like the one in \refeq{rapjetslep}. 

\begin{figure}
\centering\subfloat[$\muF=\muR=M_{\mathrm W}$]{\label{ylepFS}\epsfig{figure=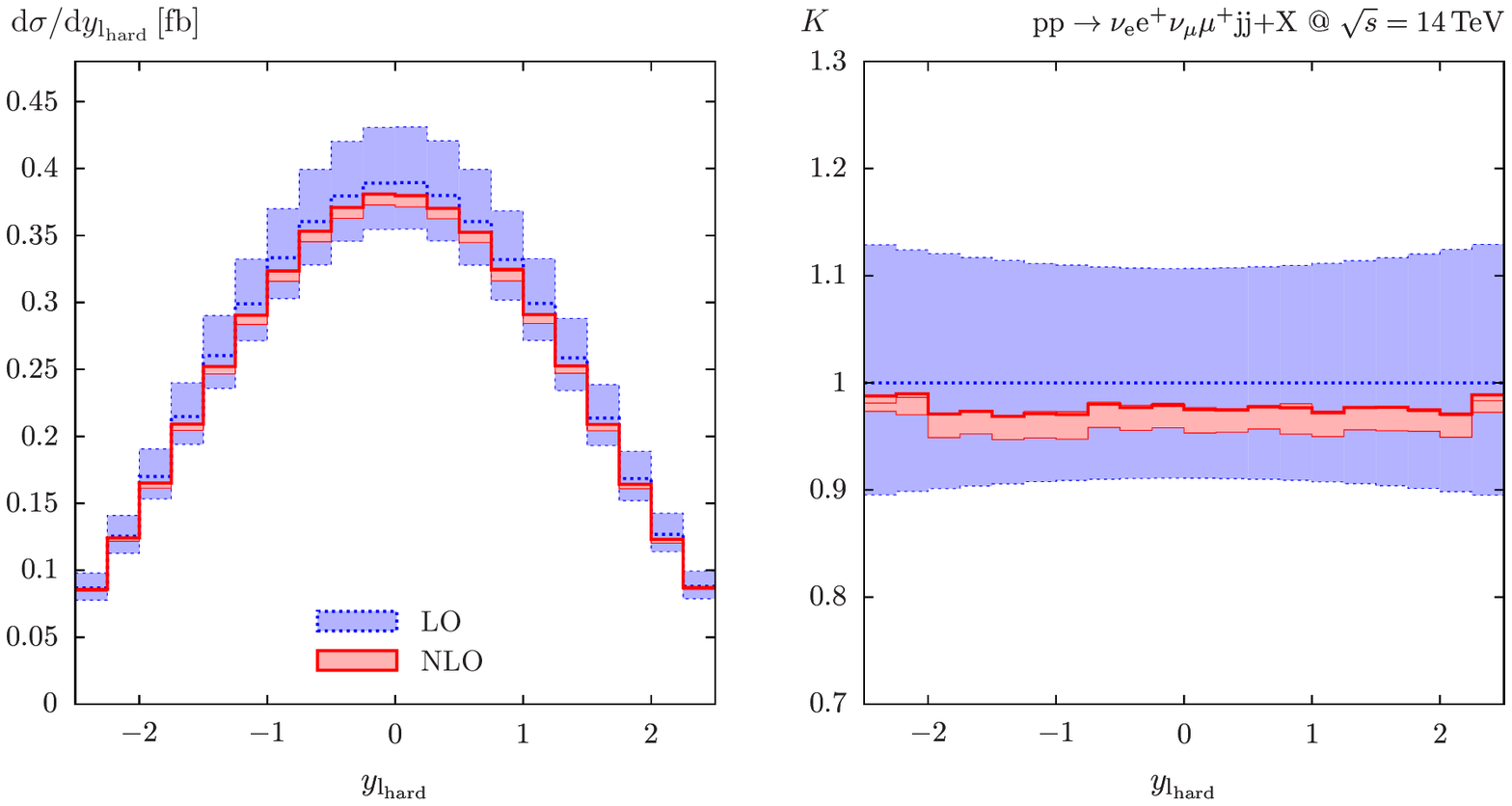,width=14cm}}\\
\subfloat[$\muF=\muR=\sqrt{p_{\mathrm{T,j_1}}\cdot p_{\mathrm{T,j_2}}}$]{\label{ylepDS2}\epsfig{figure=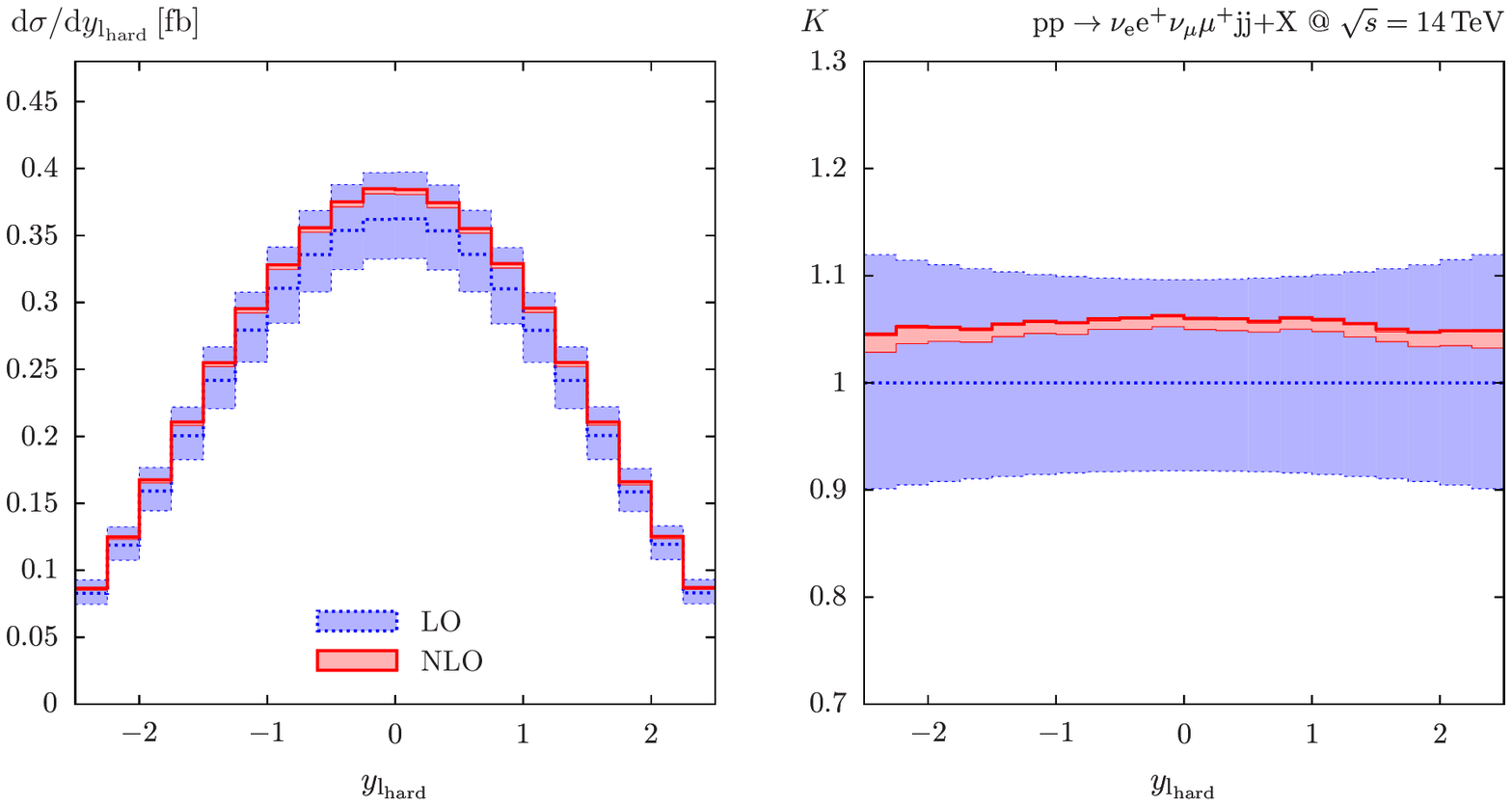,width=14cm}}
\caption{Rapidity distribution for the harder charged lepton 
for the fixed (\reffi{ylepFS}) and dynamic scale (\reffi{ylepDS2}) on the left 
and the corresponding $K$ factor represented by the solid (red) line on the right.}
\label{ylep plot}
\end{figure}

The high leptonic activity in the central region can be well observed
in the rapidity distribution of the harder (in terms of $\pT$) charged
lepton, shown in \reffi{ylep plot}.  The distribution shows a
preference for the rapidities close to zero,
while decreasing quickly as the values approach those of the tagging jets. 
One can see that for both the FS and the DS the $K$ factor remains
quite flat. As a similar behaviour can be observed in all presented
leptonic distributions, the following distributions are only shown for
the DS~(\ref{DS}).

\begin{figure}
\centering
\epsfig{figure=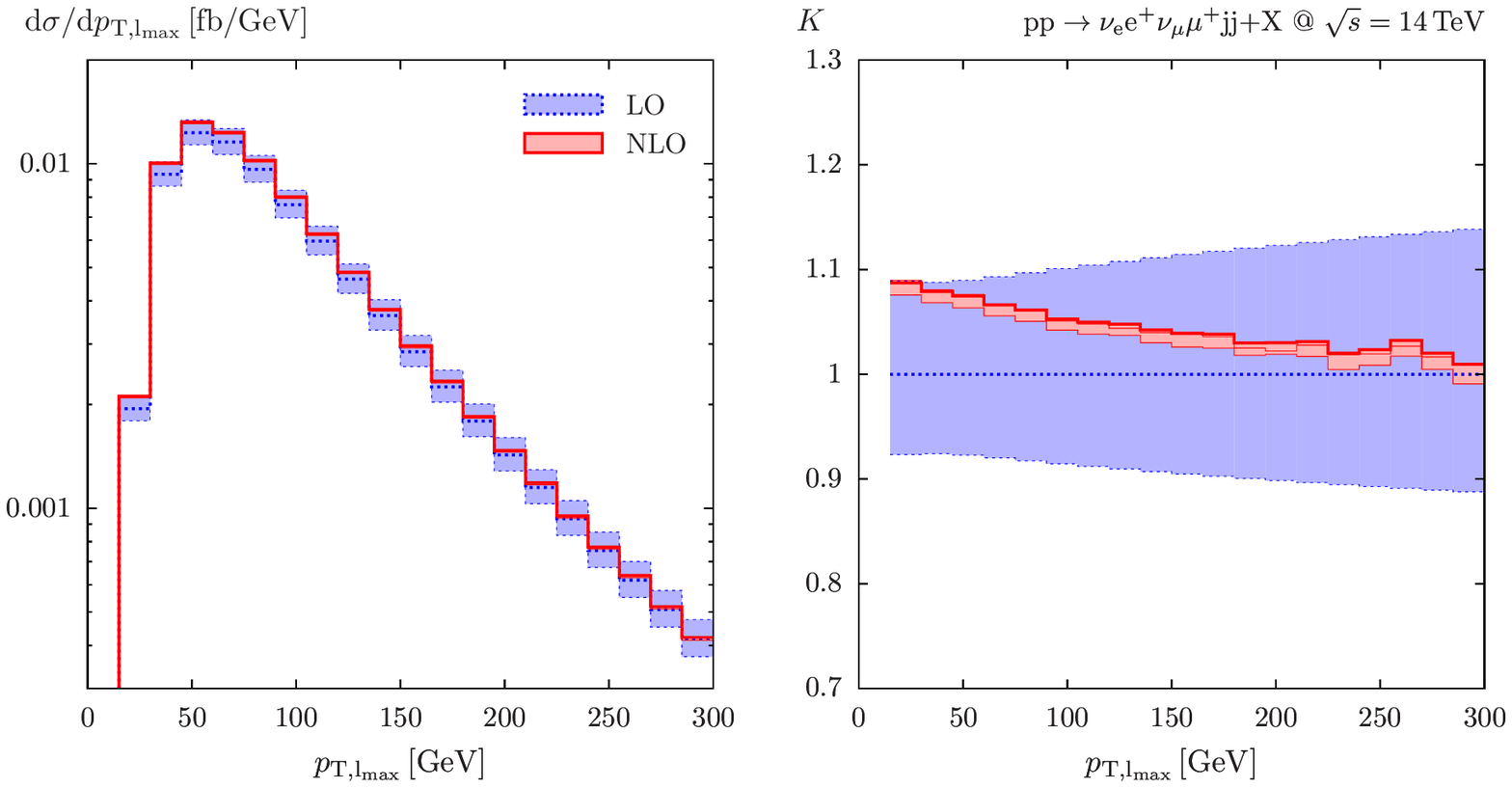,width=14cm}
\caption{Distribution of the transverse momentum of the harder charged lepton for
the dynamic scale on the left and the corresponding $K$ factor represented by
the solid (red) line on the right.}
\label{pTmaxlepDS2}
\end{figure}

\begin{figure}
\centering
\epsfig{figure=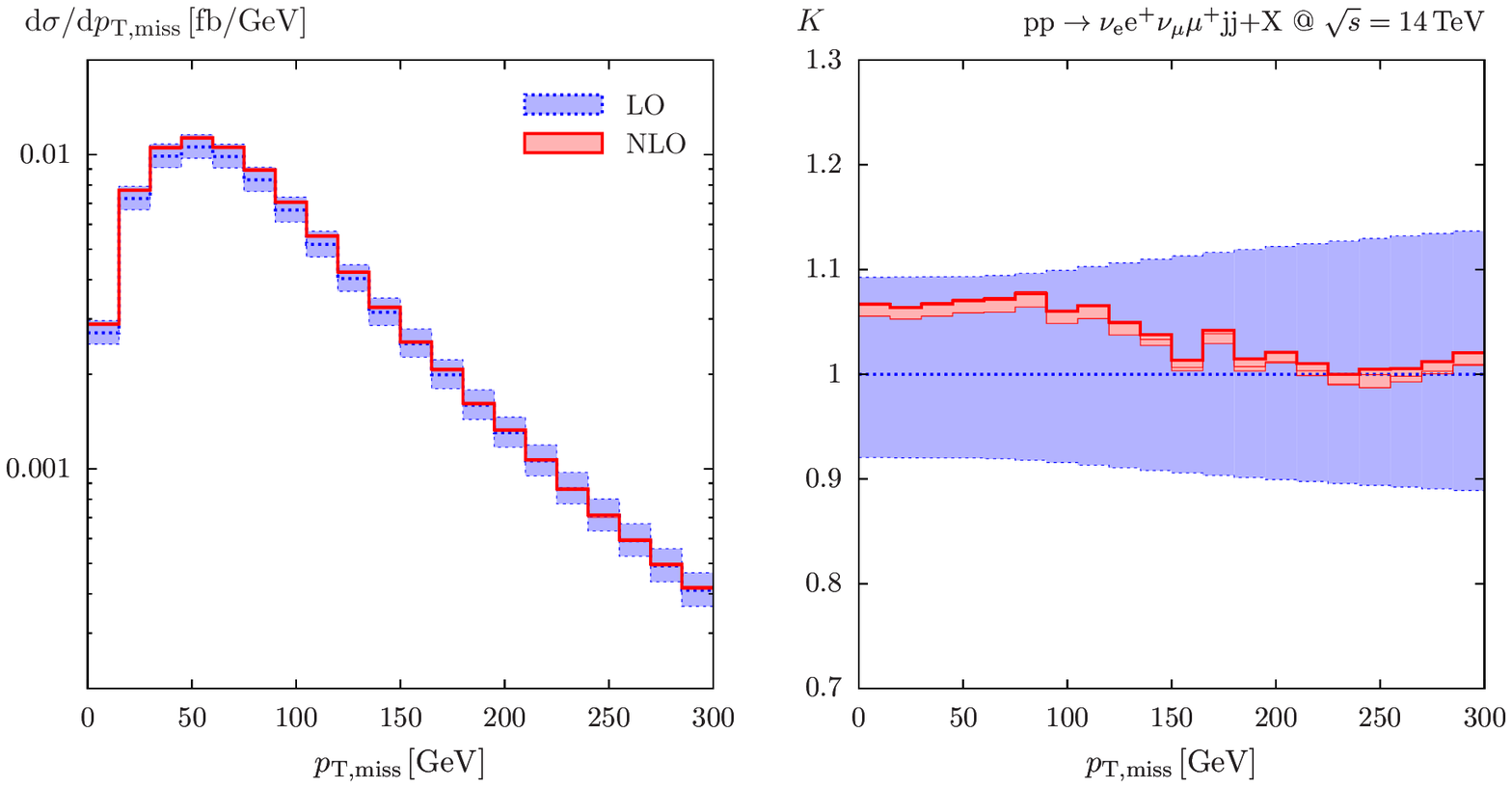,width=14cm}
\caption{Distribution of the missing transverse momentum produced by two outgoing
neutrinos for the dynamic scale on the left and the corresponding $K$ factor
represented by the solid (red) line on the right.}
\label{pTmissDS2}
\end{figure}

The distribution of the transverse momentum for the harder charged
lepton and of the missing $\pT$ corresponding to the vectorial sum of
the transverse momenta of the electron neutrino and the muon neutrino
from the W decays are shown in \reffis{pTmaxlepDS2}{pTmissDS2},
respectively.  The $K$ factors decrease with increasing transverse
momentum and are close to 1 for large $p_{\rT}$.

\begin{figure}
\centering
\epsfig{figure=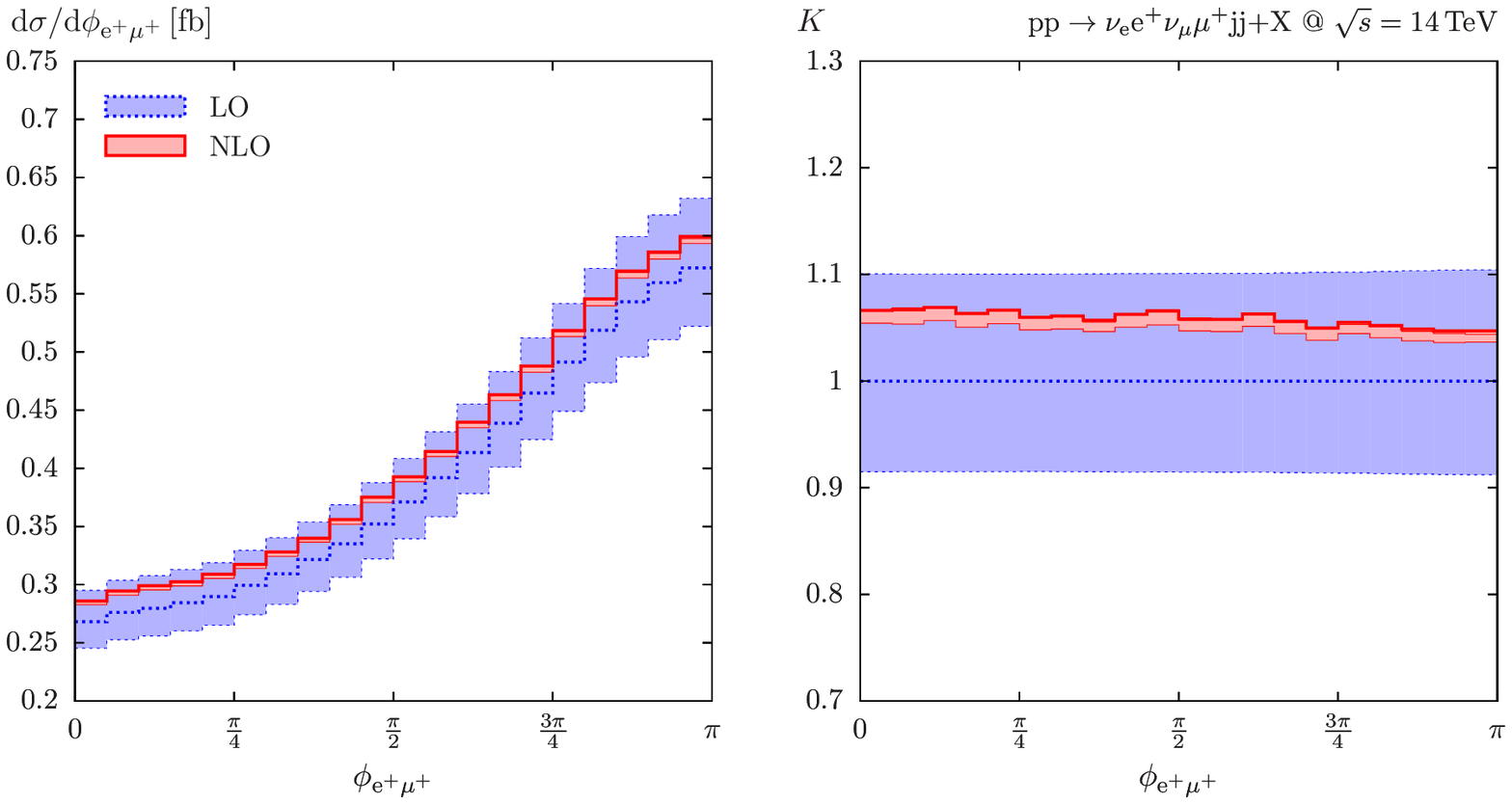,width=14cm}
\caption{Distribution of the azimuthal angle between the charged 
leptons $\re^+$ and $\mu^+$ for the dynamic scale on the left and the 
corresponding $K$ factor represented by the solid (red) line on the right.}
\label{angle plot}
\end{figure}

Angular distributions of the decay products of the vector-boson pairs
produced in VBF processes are of particular interest to the SM Higgs
searches at colliders as the leptons have a tendency to fly in the
same direction in case of a Higgs signal
\cite{Dittmar:1996sp,Rainwater:1999sd}. This is due to the fact that a
scalar particle decays into a pair of $\PWp\PWm$ which subsequently
decays into charged leptons and neutrinos, with the leptons preferably
close to each other due to the left-handed structure of the EW force.
This is not the case in the $\PWp\PWp$ production processes where a SM
Higgs decay into both intermediate vector bosons in question is
prohibited by the charges. However, a similar situation could arise in
the presence of a doubly-charged scalar resonance, which would be
produced in the VBF mode and subsequently decay into
\mbox{$\PWp\PWp\to \Pep\Pne\mu^+\nu_\mu$}.  The process
\mbox{$\Pp\Pp\to \PWp\PWp \rj\rj+X\to \Pep\Pne\mu^+\nu_\mu \rj\rj+X$}
would then deliver the dominating irreducible background,
and the angular distributions of the decay products should be well known 
to allow for an efficient background suppression.  
As demonstrated in \reffi{angle plot}, where the azimuthal angle 
$\phi_{\mathrm{\re^+\mu^+}}$
separating the charged leptons $\re^+$ and $\mu^+$ in the plane
transverse to the beam direction is depicted, the final-state leptons
are located preferentially in opposite directions in the azimuthal
plane. Similarly as with other leptonic observables, the NLO
corrections have only a modest effect in this distribution, as
exhibited by a flat $K$ factor.

\section{Conclusions}\label{chaptsummary}

This article presents a method for evaluating the NLO QCD corrections
to the electroweak production mode of processes of the form
\mbox{$\Pp\Pp\to 4l\rj\rj+X$}, associated with vector-boson
fusion of an intermediate vector-boson pair, including both resonant
contributions where the final-state leptons are produced via
vector-boson decay as well as non-resonant ones.  The Feynman diagrams
are divided into independent building blocks using internal
polarization sums and evaluated with the \textsc{FeynArts} +
\textsc{FormCalc} package in \textsc{Mathematica} using the Weyl--van
der Waerden helicity formalism. The block structure separates the
electroweak and QCD sectors of the diagrams, allowing one to apply the
QCD corrections only to building blocks involving quark lines while
electroweak building blocks are evaluated merely at tree level,
improving thus speed of the resulting \textsc{Fortran} code. The
phase-space integration is performed by a multichannel Monte Carlo
generator implemented in \textsc{C{++}} which allows to calculate
arbitrary distributions.

The described method is applied to the process
\mbox{$\Pp\Pp\to \re^+\nu_{\re}\mu^+\nu_\mu\rj\rj+X$}, which is
interesting in its own right as well as constitutes a background to
many collider searches both within and beyond the Standard Model.  The
numerical analysis is performed using typical vector-boson-fusion cuts
chosen to enhance contributions of the vector-boson-fusion
kinematics and to suppress QCD background.  The impact of the
$s$-channel diagrams and interferences between $t$ and $u$~channels is
analyzed at LO and found to be entirely negligible within
vector-boson-fusion cuts.  The NLO corrections turn out to be around
5\% of the LO cross section.  The renormalization- and
factorization-scale dependence of the cross section at NLO is reduced,
amounting to about $\pm2\%$ for the fixed scale 
\mbox{$\mu=\xi M_{\mathrm W}$} and $\pm1\%$ for the dynamical scale
\mbox{$\mu=\xi\sqrt{p_{\mathrm{T,j_1}}\cdot p_{\mathrm{T,j_2}}}$},
when varying $\xi$ within 1/2 and 2 about \mbox{$\xi=1$}, while the LO
results change significantly, by $\pm10\%$, in the same range.
Furthermore, a set of kinematical distributions for jets and
final-state leptons has been presented, demonstrating the effects of
the NLO corrections and the impact of the two scale choices. While the
fixed scale results in a strongly decreasing $K$ factor in the
high-energy tails of the distributions, the $K$ factor for the
dynamical scale approaches one in these kinematical regions.  Using
the dynamical scale at leading order provides an approximation to the
next-to-leading-order result with an accuracy below about 10\% for all
considered distributions.

\section*{Acknowledgements}

L.H. acknowledges the hospitality of the theory groups of PSI and the
University of Zurich where most of this work has been done.
This work is supported in part by the European Commission through the
Marie-Curie Research Training Network HEPTOOLS under contract
MRTN-CT-2006-035505
and through the ``LHCPhenoNet'' Initial Training Network
PITN-GA-2010-264564. We are grateful to F. Cascioli, Ph.~Maierh\"ofer,
and S. Pozzorini for providing us with \textsc{OpenLoops} matrix
elements for LO processes. We thank B. J\"ager for helping us to
compare our calculation with those of \citeres{Jager:2009xx,Jager:2011ms}.

\bibliography{ppWpWpjj}

\providecommand{\href}[2]{#2}\begingroup\raggedright\begin{thebibliography}{10}

\bibitem{Duhrssen:2004cv}
M.~{D\"uhrssen}, S.~Heinemeyer, H.~Logan, D.~Rainwater, G.~Weiglein, et~al.,
  {\it {Extracting Higgs boson couplings from CERN LHC data}},  {\em Phys.Rev.}
  {\bf D70} (2004) 113009, [\href{http://xxx.lanl.gov/abs/hep-ph/0406323}{{\tt
  hep-ph/0406323}}].

\bibitem{Belyaev:2002ua}
A.~Belyaev and L.~Reina, {\it {$pp \rightarrow t \bar{t} H$, $H \rightarrow
  \tau^+ \tau^-$: Toward a model independent determination of the Higgs boson
  couplings at the LHC}},  {\em JHEP} {\bf 0208} (2002) 041,
  [\href{http://xxx.lanl.gov/abs/hep-ph/0205270}{{\tt hep-ph/0205270}}].

\bibitem{Hankele:2006ma}
V.~Hankele, G.~{Kl\"amke}, D.~Zeppenfeld, and T.~Figy, {\it {Anomalous Higgs
  boson couplings in vector boson fusion at the CERN LHC}},  {\em Phys.Rev.}
  {\bf D74} (2006) 095001, [\href{http://xxx.lanl.gov/abs/hep-ph/0609075}{{\tt
  hep-ph/0609075}}].

\bibitem{Hagiwara:2009wt}
K.~Hagiwara, Q.~Li, and K.~Mawatari, {\it {Jet angular correlation in
  vector-boson fusion processes at hadron colliders}},  {\em JHEP} {\bf 0907}
  (2009) 101, [\href{http://xxx.lanl.gov/abs/0905.4314}{{\tt
  arXiv:0905.4314}}].

\bibitem{Bagger:1995mk}
J.~Bagger, V.~D. Barger, K.-m. Cheung, J.~F. Gunion, T.~Han, et~al., {\it {CERN
  LHC analysis of the strongly interacting W W system: Gold plated modes}},
  {\em Phys.Rev.} {\bf D52} (1995) 3878--3889,
  [\href{http://xxx.lanl.gov/abs/hep-ph/9504426}{{\tt hep-ph/9504426}}].

\bibitem{Ballestrero:2010vp}
A.~Ballestrero, D.~B. Franzosi, and E.~Maina, {\it {Vector-Vector scattering at
  the LHC with two charged leptons and two neutrinos in the final state}},
  {\em JHEP} {\bf 1106} (2011) 013,
  [\href{http://xxx.lanl.gov/abs/1011.1514}{{\tt arXiv:1011.1514}}].

\bibitem{Rainwater:1996ud}
D.~L. Rainwater, R.~Szalapski, and D.~Zeppenfeld, {\it {Probing color singlet
  exchange in $Z$ + two jet events at the CERN LHC}},  {\em Phys.Rev.} {\bf
  D54} (1996) 6680--6689, [\href{http://xxx.lanl.gov/abs/hep-ph/9605444}{{\tt
  hep-ph/9605444}}].

\bibitem{Jager:2011ms}
B.~{J\"ager} and G.~Zanderighi, {\it {NLO corrections to electroweak and QCD
  production of $W^+W^+$ plus two jets in the POWHEG BOX}},  {\em JHEP} {\bf
  1111} (2011) 055, [\href{http://xxx.lanl.gov/abs/1108.0864}{{\tt
  arXiv:1108.0864}}].

\bibitem{Dreiner:2006sv}
H.~Dreiner, S.~Grab, M.~{Kr\"amer}, and M.~Trenkel, {\it {Supersymmetric NLO
  QCD corrections to resonant slepton production and signals at the Tevatron
  and the CERN LHC}},  {\em Phys.Rev.} {\bf D75} (2007) 035003,
  [\href{http://xxx.lanl.gov/abs/hep-ph/0611195}{{\tt hep-ph/0611195}}].

\bibitem{Han:2009ya}
T.~Han, I.~Lewis, and T.~McElmurry, {\it {QCD Corrections to Scalar Diquark
  Production at Hadron Colliders}},  {\em JHEP} {\bf 1001} (2010) 123,
  [\href{http://xxx.lanl.gov/abs/0909.2666}{{\tt arXiv:0909.2666}}].

\bibitem{Maalampi:2002vx}
J.~Maalampi and N.~Romanenko, {\it {Single production of doubly charged Higgs
  bosons at hadron colliders}},  {\em Phys.Lett.} {\bf B532} (2002) 202--208,
  [\href{http://xxx.lanl.gov/abs/hep-ph/0201196}{{\tt hep-ph/0201196}}].

\bibitem{Kulesza:1999zh}
A.~Kulesza and W.~J. Stirling, {\it {Like sign $W$ boson production at the LHC
  as a probe of double parton scattering}},  {\em Phys.Lett.} {\bf B475} (2000)
  168--175, [\href{http://xxx.lanl.gov/abs/hep-ph/9912232}{{\tt
  hep-ph/9912232}}].

\bibitem{Maina:2009sj}
E.~Maina, {\it {Multiple Parton Interactions in $Z+4j, W^\pm W^\pm + 0/2j$ and
  $W^+ W^- + 2j$ production at the LHC}},  {\em JHEP} {\bf 0909} (2009) 081,
  [\href{http://xxx.lanl.gov/abs/0909.1586}{{\tt arXiv:0909.1586}}].

\bibitem{Gaunt:2010pi}
J.~R. Gaunt, C.-H. Kom, A.~Kulesza, and W.~J. Stirling, {\it {Same-sign W pair
  production as a probe of double parton scattering at the LHC}},  {\em
  Eur.Phys.J.} {\bf C69} (2010) 53--65,
  [\href{http://xxx.lanl.gov/abs/1003.3953}{{\tt arXiv:1003.3953}}].

\bibitem{Melia:2010bm}
T.~Melia, K.~Melnikov, R.~Rontsch, and G.~Zanderighi, {\it {Next-to-leading
  order QCD predictions for $W^+W^+jj$ production at the LHC}},  {\em JHEP}
  {\bf 1012} (2010) 053, [\href{http://xxx.lanl.gov/abs/1007.5313}{{\tt
  arXiv:1007.5313}}].

\bibitem{Melia:2011dw}
T.~Melia, K.~Melnikov, R.~Rontsch, and G.~Zanderighi, {\it {NLO QCD corrections
  for $W^+W^-$ pair production in association with two jets at hadron
  colliders}},  {\em Phys.Rev.} {\bf D83} (2011) 114043,
  [\href{http://xxx.lanl.gov/abs/1104.2327}{{\tt arXiv:1104.2327}}].

\bibitem{Greiner:2012im}
N.~Greiner, G.~Heinrich, P.~Mastrolia, G.~Ossola, T.~Reiter, et~al., {\it {NLO
  QCD corrections to the production of $W^+ W^-$ plus two jets at the LHC}},
  {\em Phys.Lett.} {\bf B713} (2012) 277--283,
  [\href{http://xxx.lanl.gov/abs/1202.6004}{{\tt arXiv:1202.6004}}].

\bibitem{Melia:2011gk}
T.~Melia, P.~Nason, R.~Rontsch, and G.~Zanderighi, {\it {$W^+W^+$ plus dijet
  production in the POWHEGBOX}},  {\em Eur.Phys.J.} {\bf C71} (2011) 1670,
  [\href{http://xxx.lanl.gov/abs/1102.4846}{{\tt arXiv:1102.4846}}].

\bibitem{Nason:2004rx}
P.~Nason, {\it {A New method for combining NLO QCD with shower Monte Carlo
  algorithms}},  {\em JHEP} {\bf 0411} (2004) 040,
  [\href{http://xxx.lanl.gov/abs/hep-ph/0409146}{{\tt hep-ph/0409146}}].

\bibitem{Frixione:2007vw}
S.~Frixione, P.~Nason, and C.~Oleari, {\it {Matching NLO QCD computations with
  Parton Shower simulations: the POWHEG method}},  {\em JHEP} {\bf 0711} (2007)
  070, [\href{http://xxx.lanl.gov/abs/0709.2092}{{\tt arXiv:0709.2092}}].

\bibitem{Jager:2006zc}
B.~{J\"ager}, C.~Oleari, and D.~Zeppenfeld, {\it {Next-to-leading order QCD
  corrections to $W^+ W^-$ production via vector-boson fusion}},  {\em JHEP}
  {\bf 07} (2006) 015, [\href{http://xxx.lanl.gov/abs/hep-ph/0603177}{{\tt
  hep-ph/0603177}}].

\bibitem{Jager:2006cp}
B.~{J\"ager}, C.~Oleari, and D.~Zeppenfeld, {\it {Next-to-leading order QCD
  corrections to Z boson pair production via vector-boson fusion}},  {\em Phys.
  Rev.} {\bf D73} (2006) 113006,
  [\href{http://xxx.lanl.gov/abs/hep-ph/0604200}{{\tt hep-ph/0604200}}].

\bibitem{Bozzi:2007ur}
G.~Bozzi, B.~{J\"ager}, C.~Oleari, and D.~Zeppenfeld, {\it {Next-to-leading
  order QCD corrections to $W^+Z$ and $W^-Z$ production via vector-boson
  fusion}},  {\em Phys. Rev.} {\bf D75} (2007) 073004,
  [\href{http://xxx.lanl.gov/abs/hep-ph/0701105}{{\tt hep-ph/0701105}}].

\bibitem{Jager:2009xx}
B.~{J\"ager}, C.~Oleari, and D.~Zeppenfeld, {\it {Next-to-leading order QCD
  corrections to $W^+W^+jj$ and $W^-W^-jj$ production via weak-boson fusion}},
  {\em Phys. Rev.} {\bf D80} (2009) 034022,
  [\href{http://xxx.lanl.gov/abs/0907.0580}{{\tt arXiv:0907.0580}}].

\bibitem{Lucia}
{L. Ho\v sekov\'a}, ``{\it NLO QCD corrections to the production of two lepton
  pairs via vector-boson fusion at the LHC}.'' {doctoral thesis, Zurich
  University, 2012}.

\bibitem{Ciccolini:2007ec}
M.~Ciccolini, A.~Denner, and S.~Dittmaier, {\it {Electroweak and QCD
  corrections to Higgs production via vector-boson fusion at the LHC}},  {\em
  Phys.Rev.} {\bf D77} (2008) 013002,
  [\href{http://xxx.lanl.gov/abs/0710.4749}{{\tt arXiv:0710.4749}}].

\bibitem{Hahn:2009bf}
T.~Hahn, {\it {FormCalc 6}},  {\em PoS} {\bf ACAT08} (2008) 121,
  [\href{http://xxx.lanl.gov/abs/0901.1528}{{\tt arXiv:0901.1528}}].

\bibitem{Dittmaier:1998nn}
S.~Dittmaier, {\it {Weyl-van der Waerden formalism for helicity amplitudes of
  massive particles}},  {\em Phys.Rev.} {\bf D59} (1998) 016007,
  [\href{http://xxx.lanl.gov/abs/hep-ph/9805445}{{\tt hep-ph/9805445}}].

\bibitem{Catani:1996vz}
S.~Catani and M.~H. Seymour, {\it {A general algorithm for calculating jet
  cross sections in NLO QCD}},  {\em Nucl. Phys.} {\bf B485} (1997) 291--419,
  [\href{http://xxx.lanl.gov/abs/hep-ph/9605323}{{\tt hep-ph/9605323}}].

\bibitem{Denner:2002ii}
A.~Denner and S.~Dittmaier, {\it {Reduction of one-loop tensor 5-point
  integrals}},  {\em Nucl. Phys.} {\bf B658} (2003) 175--202,
  [\href{http://xxx.lanl.gov/abs/hep-ph/0212259}{{\tt hep-ph/0212259}}].

\bibitem{Denner:2005nn}
A.~Denner and S.~Dittmaier, {\it {Reduction schemes for one-loop tensor
  integrals}},  {\em Nucl. Phys.} {\bf B734} (2006) 62--115,
  [\href{http://xxx.lanl.gov/abs/hep-ph/0509141}{{\tt hep-ph/0509141}}].

\bibitem{Denner:1991qq}
A.~Denner, U.~Nierste, and R.~Scharf, {\it {A Compact expression for the scalar
  one loop four point function}},  {\em Nucl. Phys.} {\bf B367} (1991)
  637--656.

\bibitem{Beenakker:1988jr}
W.~Beenakker and A.~Denner, {\it {Infrared Divergent Scalar Box Integrals with
  Applications in the Electroweak Standard Model}},  {\em Nucl. Phys.} {\bf
  B338} (1990) 349--370.

\bibitem{Denner:2010tr}
A.~Denner and S.~Dittmaier, {\it {Scalar one-loop 4-point integrals}},  {\em
  Nucl.Phys.} {\bf B844} (2011) 199--242,
  [\href{http://xxx.lanl.gov/abs/1005.2076}{{\tt arXiv:1005.2076}}].

\bibitem{Coli6}
A.~Denner, S.~Dittmaier, and L.~Hofer, ``{\sl COLLIER, Complex One-Loop Library
  In Extended Regularizations}.'' in preparation.

\bibitem{Denner:1999gp}
A.~Denner, S.~Dittmaier, M.~Roth, and D.~Wackeroth, {\it {Predictions for all
  processes $e^+ e^- \rightarrow$ 4 fermions ${}+ \gamma$}},  {\em Nucl.Phys.}
  {\bf B560} (1999) 33--65, [\href{http://xxx.lanl.gov/abs/hep-ph/9904472}{{\tt
  hep-ph/9904472}}].

\bibitem{Denner:2005fg}
A.~Denner, S.~Dittmaier, M.~Roth, and L.~Wieders, {\it {Electroweak corrections
  to charged-current $e^+ e^-\rightarrow$ 4 fermion processes: Technical
  details and further results}},  {\em Nucl.Phys.} {\bf B724} (2005) 247--294,
  [\href{http://xxx.lanl.gov/abs/hep-ph/0505042}{{\tt hep-ph/0505042}}].

\bibitem{Denner:2006ic}
A.~Denner and S.~Dittmaier, {\it {The complex-mass scheme for perturbative
  calculations with unstable particles}},  {\em Nucl.Phys.Proc.Suppl.} {\bf
  160} (2006) 22--26, [\href{http://xxx.lanl.gov/abs/hep-ph/0605312}{{\tt
  hep-ph/0605312}}].

\bibitem{Alwall:2007st}
J.~Alwall, P.~Demin, S.~de~Visscher, R.~Frederix, M.~Herquet, et~al., {\it
  {MadGraph/MadEvent v4: The New Web Generation}},  {\em JHEP} {\bf 0709}
  (2007) 028, [\href{http://xxx.lanl.gov/abs/0706.2334}{{\tt 0706.2334}}].

\bibitem{Cascioli:2011va}
F.~Cascioli, P.~Maierhofer, and S.~Pozzorini, {\it {Scattering Amplitudes with
  Open Loops}},  {\em Phys.Rev.Lett.} {\bf 108} (2012) 111601,
  [\href{http://xxx.lanl.gov/abs/1111.5206}{{\tt arXiv:1111.5206}}].

\bibitem{Denner:2012yc}
A.~Denner, S.~Dittmaier, S.~Kallweit, and S.~Pozzorini, {\it {NLO QCD
  corrections to off-shell top-antitop production with leptonic decays at
  hadron colliders}},  \href{http://xxx.lanl.gov/abs/1207.5018}{{\tt
  arXiv:1207.5018}}.

\bibitem{Beringer:1900zz}
{\bf Particle Data Group} Collaboration, J.~Beringer et~al., {\it {Review of
  Particle Physics (RPP)}},  {\em Phys.Rev.} {\bf D86} (2012) 010001.

\bibitem{Dittmaier:2001ay}
S.~Dittmaier and M.~{Kr\"amer}, {\it {Electroweak radiative corrections to W
  boson production at hadron colliders}},  {\em Phys.Rev.} {\bf D65} (2002)
  073007, [\href{http://xxx.lanl.gov/abs/hep-ph/0109062}{{\tt
  hep-ph/0109062}}].

\bibitem{Martin:2009iq}
A.~Martin, W.~Stirling, R.~Thorne, and G.~Watt, {\it {Parton distributions for
  the LHC}},  {\em Eur.Phys.J.} {\bf C63} (2009) 189--285,
  [\href{http://xxx.lanl.gov/abs/0901.0002}{{\tt arXiv:0901.0002}}].

\bibitem{Dittmaier:2012vm}
S.~Dittmaier, C.~Mariotti, G.~Passarino, R.~Tanaka, et~al., {\it {Handbook of
  LHC Higgs Cross Sections: 2. Differential Distributions}},
  \href{http://xxx.lanl.gov/abs/1201.3084}{{\tt arXiv:1201.3084}}.

\bibitem{Catani:1992zp}
S.~Catani, Y.~L. Dokshitzer, and B.~Webber, {\it {The $K^-$ perpendicular
  clustering algorithm for jets in deep inelastic scattering and hadron
  collisions}},  {\em Phys.Lett.} {\bf B285} (1992) 291--299.

\bibitem{Blazey:2000qt}
G.~C. Blazey, J.~R. Dittmann, S.~D. Ellis, V.~D. Elvira, K.~Frame, et~al., {\it
  {Run II jet physics}},  \href{http://xxx.lanl.gov/abs/hep-ex/0005012}{{\tt
  hep-ex/0005012}}.

\bibitem{Dittmar:1996sp}
M.~Dittmar and H.~K. Dreiner, {\it {$h^0\rightarrow W^+ W^- \rightarrow l^+ l^-
  \nu_l \bar{\nu}_l$ as the dominant SM Higgs search mode at the LHC for
  $M(h^0) = 155$GeV--$180$GeV}},
  \href{http://xxx.lanl.gov/abs/hep-ph/9703401}{{\tt hep-ph/9703401}}.

\bibitem{Rainwater:1999sd}
D.~L. Rainwater and D.~Zeppenfeld, {\it {Observing $H \rightarrow W^{(*)}
  W^{(*)} \rightarrow e^{\pm}\mu^{\mp} \slashed{p}_T$ in weak boson fusion with
  dual forward jet tagging at the CERN LHC}},  {\em Phys.Rev.} {\bf D60} (1999)
  113004, [\href{http://xxx.lanl.gov/abs/hep-ph/9906218}{{\tt
  hep-ph/9906218}}].

\end{thebibliography}\endgroup

\end{document}